# Oxides in an oxygen potential gradient: coupled morphological stability of the multiple phase boundaries

Petro O. Mchedlov-Petrosyan[1] and Manfred Martin[2]

[1] *A.I. Akhiezer Institute for Theoretical Physics, National Science Center "Kharkov Institite for Physics & Technology", 1, Akademicheskaya Str., 61108 Kharkiv, Ukraine*

[2] *Institute of Physical Chemistry, RWTH Aachen University, Landoltweg 2, 52056 Aachen Germany*

**Abstract**

In materials that are exposed to thermodynamic potential gradients, i.e., gradients of chemical potentials, electrical potential, temperature, or pressure, transport processes of the mobile components occur. These transport processes and the coupling between different processes are not only of fundamental interest, but are also the origin of several degradation processes, such as kinetic unmixing and decomposition. In addition, changes in the morphology of the material surfaces and interfaces may appear. In this paper, a comprehensive formal treatment of the coupled morphological stability of multiple phase boundaries will be given for oxides that are exposed to an oxygen potential gradient.

# I. INTRODUCTION

In many applications originally homogeneous materials are exposed to thermodynamic potential gradients, which can be gradients of temperature, chemical potential of one or more elements, electrical potential or uniaxial pressure. Well known examples are tarnishing layers on metallic materials [1,2] which act as corrosion protection, thermal barrier coatings [3] acting as heat shield, solid electrolytes in fuel cells [4], or gas separation membranes [5]. The applied gradients act as a generalized thermodynamic force, and induce directed fluxes of the mobile components. These fluxes may lead to three basic degradation phenomena of the materials. (i) The original morphology of the material surfaces might become unstable and a new surface morphology might be established (morphological instability). (ii) The multicomponent material, which was originally chemically homogeneous, becomes chemically inhomogeneous (so-called kinetic unmixing) [6]. (iii) If unmixing reaches a critical value, formation of new phases might take place, i.e., the initially single phase material might decompose into new phases (thermodynamic and/or kinetic decomposition).

The class of materials considered here will be limited to oxides. Due to their physical properties oxides are used in many technical applications, which have been discussed above. Examples are $Al_2O_3$ tarnishing layers on metallic alloys [1], $ZrO_2$-layers in thermal barrier coatings [3], $Y_2O_3$-doped $ZrO_2$ (YSZ) being the solid electrolyte in solid oxide fuel cells (SOFC) and solid oxide electrolyzer cells (SOEC), $(La,Sr)MnO_{3-d}$ being the cathode material in (SOFCs) [4], or $(La,Sr)CrO_{3-d}$ in oxygen separation membranes [5]. Recently, very thin oxide films, e.g. $SrTiO_3$ or $GaO_x$ have found increased interest due to their ability for resistive switching [7, 8]. In all of these examples, oxygen potential gradients appear across the oxide layer, either directly applied externally or as a result of another applied gradient.

In this paper we consider the most simple situation of a semiconducting binary oxide $A_{m-\delta}O_n$ where oxygen is practically immobile while cations A are mobile via cation vacancies V (with cation fraction $\delta \ll 1$). Examples are the binary transition metal oxides, $A_{m-\delta}O_n$ (A=Ni, Co, Fe, Mn). In this paper the morphological stability of AO exposed to an external oxygen potential gradient will be investigated in an exact way and will be compared to our earlier, approximate solution [9]. The results may easily be transferred to oxides where oxygen is also mobile (see [6]).

In a nonstoichiometric binary transition metal oxide $A_{1-\delta}O$ the concentration of cation vacancies increases with increasing oxygen partial pressure (or increasing temperature). If such an oxide is chemically reduced either by lowering the oxygen partial pressure (or by decreasing the temperature), then cation vacancies, V, and electron holes, $h^\bullet$, diffuse to the crystal surface, where reduction of the oxide takes place:

$$O^{2-} + V + 2h^\bullet \rightarrow \frac{1}{2} O_2(g) \qquad (1.1)$$

This reduction process corresponds to the arrival of a vacancy and two electron holes at the surface and the release of oxygen from the crystal. Thus a structural unit composed of a cation vacancy and an anion, is removed from the crystal while the number of cations is conserved. The crystal surface acts as vacancy sink until the new equilibrium state is reached. In contrast to this non-stationary situation a stationary non-equilibrium state can be established by exposing two parallel crystal surfaces of a sample to a gradient of the oxygen partial pressure, resulting in reduction at the low oxygen potential side and oxidation (the reversal of the above reaction) at the high oxygen potential side (fig. 1).

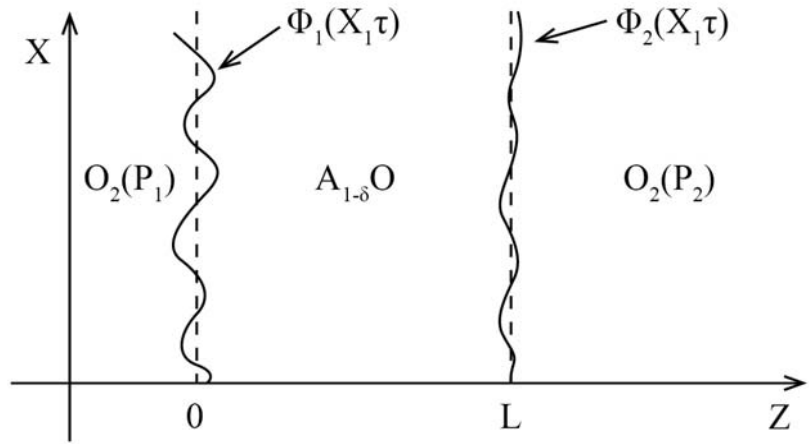

FIG. 1. Schematic presentation of on oxide $A_{1-\delta}O$ exposed to an oxygen potential gradient. $P_1$ and $P_2$ ($>P_1$) are the corresponding oxygen partial pressures in the gas phases. Dashed lines represent planar crystal surfaces, solid lines perturbed surfaces. $L$ is the width of the crystal layer.

After a transient time, a stationary flux of vacancies and a corresponding flux of A-ions in the opposite direction occur, which are fed by the interface reaction (1.1) and the reverse of it. As a result of this "vacancy wind" both crystal surfaces move (relative to the immobile oxygen sublattice) towards the side of higher partial pressure.

The corresponding one-dimensional diffusion problem can be solved easily, provided the following assumptions are made: i) the crystal surfaces are assumed to be planar. ii) the chemical diffusion coefficient $D$ describing the diffusion processes in the binary oxide is constant. iii) local equilibrium is established at the boundaries, i.e. phase boundary reaction kinetics are very fast compared to bulk diffusion. Then we can calculate a stationary solution by transforming the diffusion equation for the vacancies (or the cations A) and the mass balances at the oxide/gas boundaries to a moving reference frame, $0 \leq z \leq L$ ($L$ is the sample thickness), in which both interfaces are at rest [9]. Now, the question about the morphological stability of the surfaces of such a moving oxide layer naturally arises. Remarkably, despite the huge amount of publications on the subject of morphological stability, to the best of our knowledge the stability of the surfaces coupled by the diffusional mass transfer was not studied. In the present work the problem of linear stability of such surfaces is treated analytically. For two coupled surfaces (single layer) the results are exact; for three surfaces (two layers of different oxides of the same metal) the problem is solved in a quasistationary approximation, which is shown to be quite precise.

## II. SETTING OF THE PROBLEM

In Ref. [9] we found that the interfaces (1) and (2) exhibit different morphological stabilities. While interface (1) where the reduction takes place was morphologically unstable, interface (2) where oxidation takes place was morphologically stable. These experimental results were supported by theoretical investigations that were obtained by means of a linear stability analysis of each interface without any diffusional coupling of the interfaces. In the present work the problem of [9] is generalized in two ways: first, in exploring the morphological stability of two crystal surfaces their interaction is taken into account; second we consider also two oxide layers, i.e. two gas/solid surfaces and one solid/solid interface.

It is worth mentioning that from the formal point the result of [9] may look paradoxical: if the interaction of the boundaries is taken into account the perturbations of the boundaries are governed by a coupled *linear* equation system; a linear system could be either stable, or unstable *as a whole*; so formally both boundaries with necessity should be either stable or unstable. Below we see how this paradox is resolved in a very clear and physical way. To explore the stability of diffusionally interacting boundaries the method developed in [10, 11] is applied. The problem is solved analytically; the detailed descriptions of the (necessary) quite lengthy calculations can be found in the Appendixes.

### A. One oxide layer

For convenience we reiterate the problem setting from [9] (the present notations are slightly different). In the moving reference frame, moving with a constant velocity $V$ relative to the immobile oxygen sublattice (identical to the laboratory frame, see Fig. 1) the governing equation takes the form

$$\frac{\partial C}{\partial \tau} - V \frac{\partial C}{\partial Z} - D\left(\frac{\partial^2 C}{\partial X^2} + \frac{\partial^2 C}{\partial Z^2}\right) = 0. \qquad (2.1)$$

where $C = \delta/\omega$ is the vacancy concentration, $D$ the chemical diffusion coefficient, $\omega$ the molar volume of the oxide AO which is presumed to be constant that is independent on $\delta$, $X$ and $Z$ the spacial coordinates (see Fig. 1), and $\tau$ the time.

Looking for small deviations $u(X,Z,\tau)$ from the *stationary* solution $C_s(Z)$ corresponding to constant width $L$ of the oxide layer

$$C = C_s(Z) + u(X, Z, \tau) \tag{2.2}$$

and slightly non planar, non-stationary boundaries (see Fig. 1)

$$Z_1(X,\tau) = 0 + \Phi_1(X,\tau), \quad Z_2(X,\tau) = L + \Phi_2(X,\tau), \tag{2.3}$$

the (equilibrium) boundary conditions are

$$C_s\big|_{Z=0} + u\big|_{Z=0} + \frac{\partial C_s}{\partial Z}\bigg|_{Z=0} \Phi_1(X,\tau) = C_1\left(1 - \tilde{\Gamma}_1 \frac{\partial^2 \Phi_1}{\partial X^2}\right), \tag{2.4}$$

$$C_s\big|_{Z=L} + u\big|_{Z=L} + \frac{\partial C_s}{\partial Z}\bigg|_{Z=L} \Phi_2(X,\tau) = C_2\left(1 + \tilde{\Gamma}_2 \frac{\partial^2 \Phi_2}{\partial X^2}\right), \tag{2.5}$$

where $\tilde{\Gamma}_{1,2}$ are the capillary lengths. The mass balance equations at both interfaces are

$$V + \dot{\Phi}_1(X,\tau) = \frac{\omega}{1-\delta_1} D \left( \frac{\partial C_s}{\partial Z}\bigg|_{Z=0} + \frac{\partial u}{\partial Z}\bigg|_{Z=0} + \frac{\partial^2 C_s}{\partial Z^2}\bigg|_{Z=0} \Phi_1(X,\tau) \right), \tag{2.6}$$

$$V + \dot{\Phi}_2(X,\tau) = \frac{\omega}{1-\delta_2} D \left( \frac{\partial C_s}{\partial Z}\bigg|_{Z=L} + \frac{\partial u}{\partial Z}\bigg|_{Z=L} + \frac{\partial^2 C_s}{\partial Z^2}\bigg|_{Z=L} \Phi_2(X,\tau) \right). \tag{2.7}$$

Where $\delta_i, i = 1, 2$ are the deviations from stoichiometry at the corresponding boundaries.

If we look for the stationary ("zero order") solution, which is only $Z$ - dependent, both the equation (2.1) and the boundary conditions (2.4)-(2.7) simplify essentially. The solution of this system is given in Appendix 1; the corresponding stationary values [9] of the layer width $L$ and the velocity $V$ are given by Eqs. (7.10) and (7.11).

### B. Two oxide layers

If the oxygen partial pressure on the right-hand side of the $A_{1-\delta}O$ layer is further increased, the formation of the next oxide, e.g. $A_{3-\delta}O_4$ becomes possible (as an example one may consider CoO and Co3O4, respectively). If it happens, an additional interphase boundary appears between the oxides (see Fig. 2).

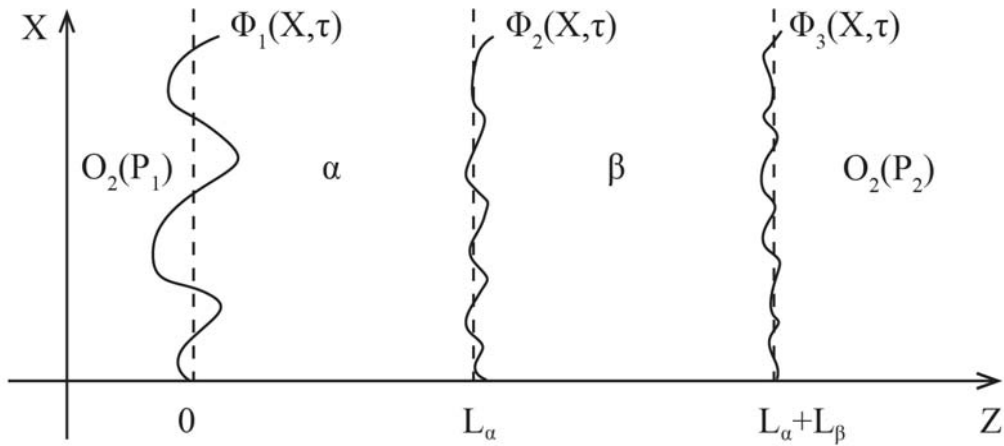

FIG. 2. Schematic presentation of two oxide layers $\alpha/\beta$, e.g. $AO/A_3O$ exposed to an oxygen potential gradient. $P_1$ and $P_2 (>P_1)$ are the corresponding oxygen partial pressures. Dashed lines represent planar crystal surfaces, solid lines perturbed surfaces.

Then, in the moving frame the governing equations take form

$$\frac{\partial C^\alpha}{\partial \tau} - V\frac{\partial C^\alpha}{\partial Z} - D_\alpha\left(\frac{\partial^2 C^\alpha}{\partial X^2} + \frac{\partial^2 C^\alpha}{\partial Z^2}\right) = 0. \tag{2.8}$$

$$\frac{\partial C^\beta}{\partial \tau} - V\frac{\partial C^\beta}{\partial Z} - D_\beta\left(\frac{\partial^2 C^\beta}{\partial X^2} + \frac{\partial^2 C^\beta}{\partial Z^2}\right) = 0. \tag{2.9}$$

where the upper index $\alpha$ refers to the AO layer and index $\beta$ refers to the $A_3O_4$ layer (see Fig. 2).

Again, for small deviations from the stationary solutions (for constant widths $L_\alpha$ and $L_\beta$),

$$C^\alpha = C_s^\alpha(Z) + u(X, Z, \tau) \tag{2.10}$$

$$C^\beta = C_s^\beta(Z) + v(X, Z, \tau) \tag{2.11}$$

and slightly non-planar, non-stationary boundaries,

$$Z_1(X, \tau) = 0 + \Phi_1(X, \tau), \tag{2.12}$$

$$Z_2(X, \tau) = L_\alpha + \Phi_2(X, \tau), \tag{2.13}$$

$$Z_3(X, \tau) = L_\alpha + L_\beta + \Phi_3(X, \tau), \tag{2.14}$$

the (equilibrium) boundary conditions are

$$C_s^\alpha\big|_{Z=0} + u\big|_{Z=0} + \frac{\partial C_s^\alpha}{\partial Z}\bigg|_{Z=0} \Phi_1(X, \tau) = C_1^\alpha \left(1 - \tilde{\Gamma}_1 \frac{\partial^2 \Phi_1}{\partial X^2}\right), \tag{2.15}$$

$$C_s^\alpha\big|_{Z=L_\alpha} + u\big|_{Z=L_\alpha} + \frac{\partial C_s^\alpha}{\partial Z}\bigg|_{Z=L_\alpha} \Phi_2(X, \tau) = C_2^\alpha \left(1 + \tilde{\Gamma}_2 \frac{\partial^2 \Phi_2}{\partial X^2}\right), \tag{2.16}$$

$$C_s^\beta\big|_{Z=L_\alpha} + v\big|_{Z=L_\alpha} + \frac{\partial C_s^\beta}{\partial Z}\bigg|_{Z=L_\alpha} \Phi_2(X, \tau) = C_2^\beta \left(1 - \tilde{\Gamma}_2 \frac{\partial^2 \Phi_2}{\partial X^2}\right), \tag{2.17}$$

$$C_s^\beta\big|_{Z=L_\alpha+L_\beta} + v\big|_{Z=L_\alpha+L_\beta} + \frac{\partial C_s^\beta}{\partial Z}\bigg|_{Z=L_\alpha+L_\beta} \Phi_3(X, t) = C_3^\beta \left(1 + \tilde{\Gamma}_3 \frac{\partial^2 \Phi_3}{\partial X^2}\right), \tag{2.18}$$

Here $\tilde{\Gamma}_i, i = 1, 2, 3$ are the capillary lengths,

$$C_i^\alpha = \frac{\delta_i^\alpha}{\omega_\alpha}, i = 1, 2 \ ; \tag{2.19}$$

$$C_i^\beta = \frac{\delta_i^\beta}{\omega_\beta}, i = 2, 3 \tag{2.20}$$

are the vacancy concentrations at the interfaces, $\delta_i^\alpha$ and $\delta_i^\beta$ are the deviations from stoichiometry at the corresponding boundaries, and $\omega_\alpha, \omega_\beta$ the molar volumes of the $\alpha$-, and $\beta$-phases. While $C_1^\alpha, C_3^\beta$ are determined by the partial pressure of oxygen at the right- and left hand sides of the oxide double layer, the concentrations $C_2^\alpha, C_2^\beta$ are determined by the local equilibrium between the adjacent oxide phases. The mass balance equations at the interfaces are

$$V + \dot{\Phi}_1(X,\tau) = \frac{\omega_\alpha}{1-\delta_1^\alpha} D_\alpha \left( \left.\frac{\partial C_s^\alpha}{\partial Z}\right|_{Z=0} + \left.\frac{\partial u}{\partial Z}\right|_{Z=0} + \left.\frac{\partial^2 C_s^\alpha}{\partial Z^2}\right|_{Z=0} \Phi_1(X,\tau) \right), \qquad (2.21)$$

$$V + \dot{\Phi}_2(X,\tau) = \left[\frac{3-\delta_2^\beta}{\omega_\beta} - \frac{1-\delta_2^\alpha}{\omega_\alpha}\right]^{-1} \left[ D_\beta \left( \left.\frac{\partial C_s^\beta}{\partial Z}\right|_{Z=L_\alpha} + \left.\frac{\partial v}{\partial Z}\right|_{Z=L_\alpha} + \left.\frac{\partial^2 C_s^\beta}{\partial Z^2}\right|_{Z=L_\alpha} \Phi_2(X,\tau) \right) - \right.$$
$$\left. - D_\alpha \left( \left.\frac{\partial C_s^\alpha}{\partial Z}\right|_{Z=L_\alpha} + \left.\frac{\partial u}{\partial Z}\right|_{Z=L_\alpha} + \left.\frac{\partial^2 C_s^\alpha}{\partial Z^2}\right|_{Z=L_\alpha} \Phi_2(X,\tau) \right) \right], \qquad (2.22)$$

$$V + \dot{\Phi}_3(X,\tau) = \frac{\omega_\beta}{3-\delta_1^\beta} D_\beta \left( \left.\frac{\partial C_s^\beta}{\partial Z}\right|_{Z=L_\alpha+L_\beta} + \left.\frac{\partial v}{\partial Z}\right|_{Z=L_\alpha+L_\beta} + \left.\frac{\partial^2 C_s^\beta}{\partial Z^2}\right|_{Z=L_\alpha+L_\beta} \Phi_3(X,\tau) \right), (2.23)$$

If we look for the stationary ("zero order") solution, quite analogous to Subsection 2.1, the equations (2.8)-(2.9), (2.15)-(2.18), and (2.21)-(2.23) simplify essentially again. The corresponding system of equations and the boundary conditions are given in Appendix 2; the stationary values of the layer widths $L_\alpha, L_\beta$ and the velocity $V$ are now

$$L_\alpha = L_0 D_\alpha \frac{1-\delta_0^\alpha}{\omega_\alpha} \left[\frac{D_\alpha}{\omega_\alpha}(\delta_2^\alpha - \delta_1^\alpha) + \frac{D_\beta}{\omega_\beta}(\delta_3^\beta - \delta_2^\beta)\right]^{-1} \ln\frac{1-\delta_1^\alpha}{1-\delta_2^\alpha}. \qquad (2.24)$$

$$L_\beta = L_0 D_\beta \frac{1-\delta_0^\alpha}{\omega_\alpha} \left[\frac{D_\alpha}{\omega_\alpha}(\delta_2^\alpha - \delta_1^\alpha) + \frac{D_\beta}{\omega_\beta}(\delta_3^\beta - \delta_2^\beta)\right]^{-1} \ln\frac{3-\delta_2^\beta}{3-\delta_3^\beta}. \qquad (2.25)$$

$$V = \frac{\omega_\alpha}{L_0\left(1-\delta_0^\alpha\right)}\left[\frac{D_\alpha}{\omega_\alpha}\left(\delta_2^\alpha - \delta_1^\alpha\right) + \frac{D_\beta}{\omega_\beta}\left(\delta_3^\beta - \delta_2^\beta\right)\right] \quad (2.26)$$

Naturally, setting $\delta_3^\beta = \delta_2^\beta$ in (2.24) we regain the expression for the stationary width of a single $\alpha$-layer (7.10), and setting $\delta_2^\alpha = \delta_1^\alpha$ in (2.25) we obtain the similar expression for the single $\beta$-layer.

### III. STABILITY ANALYSIS FOR THE SINGLE OXIDE LAYER

First we consider the single oxide layer, i.e. is the coupled stability of two gas/solid interfaces. The governing equation and boundary conditions for the first order perturbations are, see (2.1)-(2.7),

$$\frac{1}{D}\frac{\partial u}{\partial \tau} = \frac{V}{D}\frac{\partial u}{\partial Z} + \frac{\partial^2 u}{\partial X^2} + \frac{\partial^2 u}{\partial Z^2}, \quad (3.1)$$

$$u\big|_{Z=0} + \frac{V}{D}\frac{1-\delta_1}{\omega}\Phi_1(X,\tau) + C_1\tilde{\Gamma}_1\frac{\partial^2 \Phi_1}{\partial X^2} = 0, \quad (3.2)$$

$$u\big|_{Z=L} + \frac{V}{D}\frac{1-\delta_2}{\omega}\Phi_2(X,\tau) - C_2\tilde{\Gamma}_2\frac{\partial^2 \Phi_2}{\partial X^2} = 0, \quad (3.3)$$

$$\frac{1}{D}\frac{\partial \Phi_1}{\partial \tau} = \frac{\omega}{1-\delta_1}\frac{\partial u}{\partial Z}\bigg|_{Z=0} - \left(\frac{V}{D}\right)^2 \Phi_1(X,\tau), \quad (3.4)$$

$$\frac{1}{D}\frac{\partial \Phi_2}{\partial \tau} = \frac{\omega}{1-\delta_2}\frac{\partial u}{\partial Z}\bigg|_{Z=L} - \left(\frac{V}{D}\right)^2 \Phi_2(X,\tau), \quad (3.5)$$

where we have used the expressions (7.12)-(7.13) for the values of the derivatives $\frac{\partial C_s}{\partial Z}\bigg|_{Z=0}, \frac{\partial^2 C_s}{\partial Z^2}\bigg|_{Z=0}$ and $\frac{\partial C_s}{\partial Z}\bigg|_{Z=L}, \frac{\partial^2 C_s}{\partial Z^2}\bigg|_{Z=L}$, see Appendix 1. Taking the stationary width $L$ of the layer as the length scale and, correspondingly rescaling all other lengths $Z/L = z$,

$X/L = x$, $\Phi_i/L = \varphi_i$, $\tilde{\Gamma}_i/L = \Gamma_i$ and time $\tau / \left(\dfrac{L^2}{D}\right) = t$, and measuring $u$ in molar fractions $\omega u = \tilde{u}$, we are led to the dimensionless system for the concentration's perturbation field

$$\frac{\partial \tilde{u}}{\partial t} = 2\xi \frac{\partial \tilde{u}}{\partial z} + \frac{\partial^2 \tilde{u}}{\partial x^2} + \frac{\partial^2 \tilde{u}}{\partial z^2}, \qquad (3.6)$$

where

$$\xi = \frac{VL}{2D} = \frac{1}{2}\ln\frac{1-\delta_1}{1-\delta_2}, \qquad (3.7)$$

The boundary conditions at the positions of the planar boundaries, $z=0$ and $z=1$, are

$$\tilde{u}\big|_{z=0} + 2\xi(1-\delta_1)\varphi_1(x,t) + \delta_1 \Gamma_1 \frac{\partial^2 \varphi_1}{\partial x^2} = 0, \qquad (3.8)$$

$$\tilde{u}\big|_{z=1} + 2\xi(1-\delta_2)\varphi_2(x,t) - \delta_2 \Gamma_2 \frac{\partial^2 \varphi_2}{\partial x^2} = 0, \qquad (3.9)$$

$$\frac{\partial \varphi_1}{\partial t} = \frac{1}{1-\delta_1}\frac{\partial \tilde{u}}{\partial z}\bigg|_{z=0} - 4\xi^2 \varphi_1(x,t), \qquad (3.10)$$

$$\frac{\partial \varphi_2}{\partial t} = \frac{1}{1-\delta_2}\frac{\partial \tilde{u}}{\partial z}\bigg|_{z=1} - 4\xi^2 \varphi_2(x,t). \qquad (3.11)$$

Introducing the Fourier transforms,

$$\varphi_j(x,t) = \frac{1}{\sqrt{2\pi}} \int_{-\infty}^{\infty} dk \exp(ikx) \bar{\varphi}_j(k,t), \quad j=1,2 \qquad (3.12)$$

$$\tilde{u}(x,z,t) = \frac{1}{\sqrt{2\pi}} \int_{-\infty}^{\infty} dk \exp(ikx) \bar{u}(k,z,t), \qquad (3.13)$$

we obtain from (3.6) and (3.8)-(3.11)

$$\frac{\partial \bar{u}}{\partial t} = 2\xi \frac{\partial \bar{u}}{\partial z} + \frac{\partial^2 \bar{u}}{\partial z^2} - k^2 \bar{u}, \qquad (3.14)$$

$$\bar{u}|_{z=0} + 2\xi(1-\delta_1)\bar{\varphi}_1 - \delta_1\Gamma_1 k^2\bar{\varphi}_1 = 0, \tag{3.15}$$

$$\bar{u}|_{z=1} + 2\xi(1-\delta_2)\bar{\varphi}_2 + \delta_2\Gamma_2 k^2\bar{\varphi}_2 = 0, \tag{3.16}$$

$$\frac{\partial \bar{\varphi}_1}{\partial t} = \frac{1}{1-\delta_1}\frac{\partial \bar{u}}{\partial z}\bigg|_{z=0} - 4\xi^2\bar{\varphi}_1, \tag{3.17}$$

$$\frac{\partial \bar{\varphi}_2}{\partial t} = \frac{1}{1-\delta_2}\frac{\partial \bar{u}}{\partial z}\bigg|_{z=1} - 4\xi^2\bar{\varphi}_2. \tag{3.18}$$

The method [10-11] to be applied below essentially uses the asymptotical (in time) finiteness of $\bar{u}$, $\bar{\varphi}_1$ and $\bar{\varphi}_2$. Therefore, to take into account the possible instability (which we are looking for) the new variables are introduced:

$$w = \bar{u}\exp(-\eta t), \tag{3.19}$$

$$\gamma_i = \bar{\varphi}_i \exp(-\eta t), \quad i = 1,2 \tag{3.20}$$

where the constant $\eta > 0$ at the moment remains undetermined. In terms of these new variables (3.14)-(3.18) become

$$\frac{\partial w}{\partial t} = 2\xi\frac{\partial w}{\partial z} + \frac{\partial^2 w}{\partial z^2} - (k^2 + \eta)w, \tag{3.21}$$

$$w|_{z=0} + \left[2\xi(1-\delta_1) - \delta_1\Gamma_1 k^2\right]\gamma_1(k,t) = 0, \tag{3.22}$$

$$w|_{z=1} + \left[2\xi(1-\delta_2) + \delta_2\Gamma_2 k^2\right]\gamma_2(k,t) = 0, \tag{3.23}$$

$$\frac{\partial \gamma_1}{\partial t} = \frac{1}{1-\delta_1}\frac{\partial w}{\partial z}\bigg|_{z=0} - (4\xi^2 + \eta)\gamma_1, \tag{3.24}$$

$$\frac{\partial \gamma_2}{\partial t} = \frac{1}{1-\delta_2}\frac{\partial w}{\partial z}\bigg|_{z=1} - (4\xi^2 + \eta)\gamma_2. \tag{3.25}$$

The term $\eta \cdot \omega$ in (3.21) may be considered physically as some kind of fictitious "dissipation", which may be adjusted to compensate the possible instability. Now the method [10-11] is applied. First we perform the integral transformations of (3.21), using as a kernel functions

$$K_n = \exp(\nu_n z), \ n = 1, 2, \quad (3.26)$$

where

$$\nu_{1,2} = \xi \pm \sqrt{\xi^2 + k^2 + \eta + p}; \ p > 0. \quad (3.27)$$

I.e., is we introduce

$$I_n(k, p, t) = \int_0^1 w(k, z, t) \exp(\nu_n z) dz, \ n = 1, 2. \quad (3.28)$$

This yields two ordinary differential equations for $I_n$

$$\frac{\partial I_n}{\partial t} = p I_n + \Phi_n, \ n = 1, 2, \quad (3.29)$$

where

$$\Phi_n = (2\xi - \nu_n)\left[w|_{z=1} \exp(\nu_n) - w|_{z=0}\right] + \left.\frac{\partial w}{\partial z}\right|_{z=1} \exp(\nu_n) - \left.\frac{\partial w}{\partial z}\right|_{z=0}. \quad (3.30)$$

Solving (3.29) we obtain

$$I_n(p, k, t) \exp(-pt) = \int_0^t \Phi_n \exp(-pq) dq + I_{n0}, \ n = 1, 2, \quad (3.31)$$

$$I_{n0} = \int_0^1 \overline{u}(k, z, 0) \exp(\nu_n z) dz, \ n = 1, 2, \quad (3.32)$$

where $\overline{u}(k, z, 0)$ is the Fourier transform (see (3.13)) of initial deviations of the concentration from the stationary solution inside the layer.

Even if the stationary solution appears to be unstable, that is the boundary perturbations

44

$\varphi_i$ are increasing with time, selecting $\eta$ (see (3.19),(3.20)), we can always "shift" $w, \gamma_i$ to the stability threshold. At the threshold $I_n$ are (see (3.28)) limited for $t \to \infty$. Taking this limit on both sides of (3.31), we get

$$0 = \int_0^\infty \Phi_n \exp(-pq) dq + I_{n0}(k, p, 0), \quad n = 1, 2. \quad (3.33)$$

Using the boundary conditions (3.22)-(3.25), $w$ and its derivatives could be eliminated from (3.30), and $\Phi_n$ could be expressed via $\gamma_i$ and $\dfrac{\partial \gamma_i}{\partial t}$ only; this yields a system of two integral equations for $\gamma_1$ and $\gamma_2$. The method, described above, was developed in [10-11] on the basis of an approach designed by Chekmaryova [13] for the solution of the one-dimensional moving boundary problems for diffusion equations. While for the moving boundary problems the integral equations (the analogue of (3.33)) are highly nonlinear, (3.33) is a linear one. Even more, for the present case it turns out to be a linear equation for the Laplace transforms of $\gamma_1$, $\gamma_2$. Denoting the Laplace transforms of $\gamma_i$ as $\hat{\gamma}_i$, we arrive at the following algebraic system of equations for $\hat{\gamma}_i$,

$$\begin{aligned} & \left[(1 - \delta_2)(2\xi \nu_1 + p + \eta) - \nu_2 \delta_2 \Gamma_2 k^2\right] \hat{\gamma}_2 - \\ & - \exp(\nu_1)\left[(1 - \delta_1)(2\xi \nu_1 + p + \eta) + \nu_2 \delta_1 \Gamma_1 k^2\right] \hat{\gamma}_1 = \\ & = -I_{10}(k, p, 0) + (1 - \delta_2) \exp(\nu_1) \gamma_2(k, 0) - (1 - \delta_1) \gamma_1(k, 0) \end{aligned} \quad (3.34)$$

$$\begin{aligned} & \exp(\nu_2)\left[(1 - \delta_2)(2\xi \nu_2 + p + \eta) - \nu_1 \delta_2 \Gamma_2 k^2\right] \hat{\gamma}_2 - \\ & - \left[(1 - \delta_1)(2\xi \nu_2 + p + \eta) + \nu_1 \delta_1 \Gamma_1 k^2\right] \hat{\gamma}_1 = \\ & = -I_{20}(k, p, 0) + (1 - \delta_2) \exp(\nu_2) \gamma_2(k, 0) - (1 - \delta_1) \gamma_1(k, 0). \end{aligned} \quad (3.35)$$

where $\gamma_i(k, 0) = \overline{\varphi}_i(k, 0)$, that is initial values of the $k$-th Fourier mode of the boundaries'

perturbations. Introducing notations

$$\overline{\Gamma}_i = \frac{\delta_i \Gamma_i}{1-\delta_i}; i=1,2; \ \overline{I}_{10} = I_{10}\frac{1}{1-\delta_1}; \overline{I}_{20} = \frac{1}{1-\delta_2}I_{20}, \qquad (3.36)$$

and taking into account (see(3.7) and (3.27))

$$\frac{1-\delta^{(2)}}{1-\delta^{(1)}} = \exp(-2\xi); \ \nu_1 + \nu_2 = 2\xi,$$

we rewrite (3.34), (3.35), as

$$\begin{aligned}\exp(-\nu_2)\Big[(2\xi\nu_1 + p + \eta) - \nu_2\overline{\Gamma}_2 k^2\Big]\hat{\gamma}_2 - \\ -\Big[(2\xi\nu_1 + p + \eta) + \nu_2\overline{\Gamma}_1 k^2\Big]\hat{\gamma}_1 = \\ = -\overline{I}_{10}(k,p,0) + \exp(-\nu_2)\gamma_2(k,0) - \gamma_1(k,0)\end{aligned} \qquad (3.37)$$

$$\begin{aligned}\exp(-\nu_1)\Big[(2\xi\nu_2 + p + \eta) - \nu_1\overline{\Gamma}_2 k^2\Big]\hat{\gamma}_2 - \\ -\Big[(2\xi\nu_2 + p + \eta) + \nu_1\overline{\Gamma}_1 k^2\Big]\hat{\gamma}_1 = \\ = -\overline{I}_{20}(k,p,0) + \exp(-\nu_1)\gamma_2(k,0) - \gamma_1(k,0)\end{aligned} \qquad (3.38)$$

*Zero surface tension case*. In the present work we are mainly targeting the effect of the diffusional interaction of the moving boundaries on their morphological stability. Both renormalized capillary lengths $\overline{\Gamma}_i$ are quite small (see (3.36): $\Gamma_i = \frac{\tilde{\Gamma}_i}{L} \sim 10^{-6}$; $\frac{\delta_i}{1-\delta_i} \sim 10^{-2}$, or less). That is, the influence of surface tension may be essential only for the perturbations with the wave length less than $10^{-4} \times L$ [9]. So we may first set for simplicity $\overline{\Gamma}_i = 0$, postponing the discussion of the nonzero $\Gamma_i$ to a future work. We also do not consider the effect of initial perturbations of the concentration field *inside* the layer, taking $\bar{I}_{10} = \bar{I}_{20} = 0$. It was shown in [10] that for rather weak assumptions about the initial perturbations of the concentration field inside the layer (expandability in a converging Taylor series) their influence on the boundaries' instability is negligible. Then the system (3.37)-(3.38) takes the form:

$$\begin{aligned}(2\xi\nu_1 + p + \eta)\left[\exp(-\nu_2)\hat{\gamma}_2 - \hat{\gamma}_1\right] = \\ = \exp(-\nu_2)\gamma_2(k,0) - \gamma_1(k,0)\end{aligned} \quad (3.39)$$

$$\begin{aligned}(2\xi\nu_2 + p + \eta)\left[\exp(-\nu_1)\hat{\gamma}_2 - \hat{\gamma}_1\right] = \\ = \exp(-\nu_1)\gamma_2(k,0) - \gamma_1(k,0)\end{aligned} \quad (3.40)$$

Solving system (3.39)-(3.40) for $\hat{\gamma}_i$, we obtain

$$\hat{\gamma}_1 = \hat{F}_{11}\gamma_1(k,0) + \hat{F}_{12}\gamma_2(k,0), \quad (3.41)$$

$$\hat{\gamma}_2 = \hat{F}_{21}\gamma_1(k,0) + \hat{F}_{22}\gamma_2(k,0), \quad (3.42)$$

where the functions $\hat{F}_{ij}$ are:

$$\hat{F}_{11} = \frac{1}{1 - \exp(\nu_2 - \nu_1)}\left[\frac{1}{2\xi\nu_2 + p + \eta} - \frac{\exp(\nu_2 - \nu_1)}{2\xi\nu_1 + p + \eta}\right] \quad (3.43)$$

$$\hat{F}_{12} = \frac{\exp(-\nu_1)}{1 - \exp(\nu_2 - \nu_1)}\left[\frac{1}{2\xi\nu_1 + p + \eta} - \frac{1}{2\xi\nu_2 + p + \eta}\right] \quad (3.44)$$

$$\hat{F}_{21} = \frac{\exp(\nu_2)}{1 - \exp(\nu_2 - \nu_1)}\left[\frac{1}{2\xi\nu_2 + p + \eta} - \frac{1}{2\xi\nu_1 + p + \eta}\right] \quad (3.45)$$

$$\hat{F}_{22} = \frac{1}{1-\exp(\nu_2-\nu_1)}\left[\frac{1}{2\xi\nu_1+p+\eta} - \frac{\exp(\nu_2-\nu_1)}{2\xi\nu_2+p+\eta}\right] \quad (3.46)$$

Performing the inverse Laplace transform, one obtains the exact (fully time-dependent) expressions for the evolution of the boundary perturbations:

$$\gamma_1(k,t) = F_{11}\gamma_1(k,0) + F_{12}\gamma_2(k,0) \quad (3.47)$$

$$\gamma_2(k,t) = F_{21}\gamma_1(k,0) + F_{22}\gamma_2(k,0) \quad (3.48)$$

Here

$$F_{ij} = \frac{1}{2\pi i}\int_{a-i\infty}^{a+i\infty} \hat{F}_{ij}(p,k)\exp(pt)\,dp \quad (3.49)$$

It is evident from (3.47)-(3.48) that the $F_{11}$ and $F_{22}$ exhibit the "self-action" of the reducing and oxidizing boundaries, respectively, that is the evolution of their own initial perturbations. On the other hand, the $F_{12}$ and $F_{21}$ reveal the "cross-action", that is the influence of the initial perturbation of the oxidizing boundary on the evolution of the reducing boundary, and vice versa.

Of cause, this result may be obtained by the complete solution of the problem (3.21)-(3.25) via the Laplace transformation. However we succeeded in obtaining $\gamma_i(k,t)$ only, *without solving the problem completely*. With increasing number of the boundaries, and/or of the components, this difference becomes increasingly important. Even more, to explore stability we do not need the full solutions for $\gamma_i(k,t)$. It is sufficient to detect the fastest growing modes only, which, in turn are determined by the singularities of the corresponding integrands in (3.49). It is convenient to introduce the new variable $y$:

$$y = \xi^2 + k^2 + \eta + p \qquad (3.50)$$

Then $\nu_1 = \xi + \sqrt{y}$ ; $\nu_2 = \xi - \sqrt{y}$ , see Eq.(3.27); $p = y - \xi^2 - k^2 - \eta$ and (3.43)-(3.46) may be rewritten as

$$\hat{F}_{11} = \frac{1}{1 - \exp\left(-2\sqrt{y}\right)} \left[ \frac{1}{\left(\sqrt{y} - \xi - k\right)\left(\sqrt{y} - \xi + k\right)} - \frac{\exp\left(-2\sqrt{y}\right)}{\left(\sqrt{y} + \xi - k\right)\left(\sqrt{y} + \xi + k\right)} \right] \qquad (3.51)$$

$$\hat{F}_{12} = \frac{\exp\left(-\xi - \sqrt{y}\right)}{1 - \exp\left(-2\sqrt{y}\right)} \left[ \frac{1}{\left(\sqrt{y} + \xi - k\right)\left(\sqrt{y} + \xi + k\right)} - \frac{1}{\left(\sqrt{y} - \xi - k\right)\left(\sqrt{y} - \xi + k\right)} \right] \qquad (3.52)$$

$$\hat{F}_{21} = \frac{\exp\left(\xi - \sqrt{y}\right)}{1 - \exp\left(-2\sqrt{y}\right)} \left[ \frac{1}{\left(\sqrt{y} - \xi - k\right)\left(\sqrt{y} - \xi + k\right)} - \frac{1}{\left(\sqrt{y} + \xi - k\right)\left(\sqrt{y} + \xi + k\right)} \right] \qquad (3.53)$$

$$\hat{F}_{22} = \frac{1}{1 - \exp\left(-2\sqrt{y}\right)} \left[ \frac{1}{\left(\sqrt{y} + \xi - k\right)\left(\sqrt{y} + \xi + k\right)} - \frac{\exp\left(-2\sqrt{y}\right)}{\left(\sqrt{y} - \xi - k\right)\left(\sqrt{y} - \xi + k\right)} \right] \qquad (3.54)$$

Making use of (3.49) we get

$$F_{lr} = \frac{1}{2\pi i} \exp\{-(\xi^2 + k^2 + \eta)t\}\left[J_{lr}^{(1)} - J_{lr}^{(2)}\right], l, r = 1, 2. \quad (3.55)$$

From the eight integrals $J_{lr}^{(m)}, l, r, m = 1, 2$ we show here only $J_{11}^{(1)}$ as an example; all $J_{lr}^{(m)}$ are given in the Appendix 3.

$$J_{11}^{(1)} = \int_{a-i\infty}^{a+i\infty} \frac{\left(\sqrt{y} + \xi + k\right)\exp(yt)}{\left(1 - \exp(-2\sqrt{y})\right)\left[y - (k+\xi)^2\right]\left(\sqrt{y} - \xi + k\right)} dy \quad (3.56)$$

The integrand of each $J_{lr}^{(m)}$ has a branching point at $y = 0$ and two poles; only the pole with the maximal real part is of interest. Thus the 8 integrals $J_{lr}^{(m)}$ are segregated into two sets: those with maximal real part of the pole $(k + \xi)^2$, and those with maximal real part of the pole $(k - \xi)^2$. The former set includes $J_{11}^{(1)}, J_{12}^{(2)}, J_{21}^{(1)}$ and $J_{22}^{(2)}$; the latter $J_{11}^{(2)}, J_{12}^{(1)}, J_{21}^{(2)}$ and $J_{22}^{(1)}$. The integration contours for the integrals of the first and second set are shown in the Fig. 3 and Fig. 4, respectively.

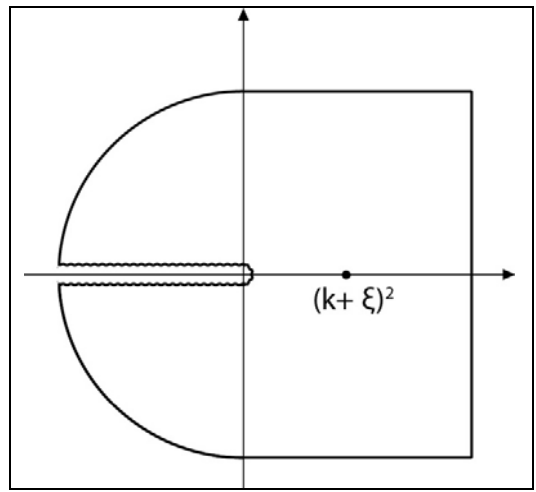

FIG. 3. Integration contours for $J_{11}^{(1)}, J_{12}^{(2)}, J_{21}^{(1)}, J_{22}^{(2)}$ in the complex plane $y$. Only position of the poles with maximal real part is shown.

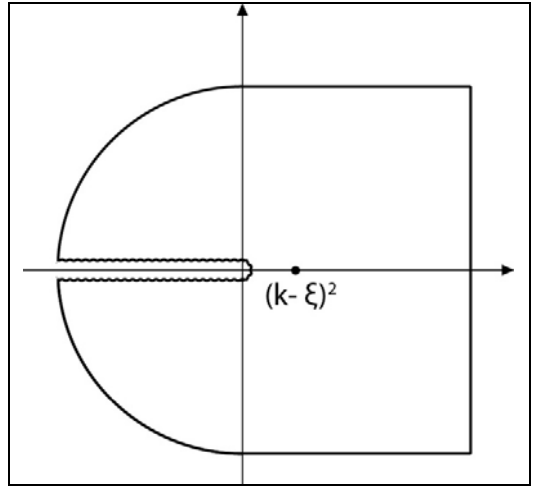

FIG. 4. Integration contours for $J_{11}^{(2)}, J_{12}^{(1)}, J_{21}^{(2)}, J_{22}^{(1)}$ in the complex plane $y$. Only position of the poles with maximal real part is shown.

Calculation of the residues at the poles and integration along the cut reveals that the fastest growing terms correspond to the input of residues at the poles of the first-set integrals. Taking into account (3.55), we get finally for the $F_{ij}$ in (3.47)-(3.48):

$$F_{11} \sim \exp\{(2\xi k - \eta)t\} \qquad (3.57)$$

$$F_{12} \sim \exp\{-(2\xi + k) + (2\xi k - \eta)t\} \qquad (3.58)$$

$$F_{21} \sim \exp\{-k + (2\xi k - \eta)t\} \qquad (3.59)$$

$$F_{22} \sim \exp\{-2(\xi + k) + (2\xi k - \eta)t\} \qquad (3.60)$$

The margin of stability for $\gamma_i$ in (3.47), (3.48) is $\eta = 2\xi k$. In terms of the Fourier modes of the boundaries' perturbations $\overline{\varphi}_i, i = 1, 2$, see (3.20), this means the increment $2\xi k$ for the $k$-th mode at both boundaries, i.e.

$$\overline{\varphi}_1(k,t) = \tilde{F}_{11}\overline{\varphi}_1(k,0) + \tilde{F}_{12}\overline{\varphi}_2(k,0) \tag{3.61}$$

$$\overline{\varphi}_2(k,t) = \tilde{F}_{21}\overline{\varphi}_1(k,0) + \tilde{F}_{22}\overline{\varphi}_2(k,0), \tag{3.62}$$

where

$$\tilde{F}_{11} \sim \exp(2\xi k t) \tag{3.63}$$

$$\tilde{F}_{12} \sim \exp\{-(2\xi + k) + 2\xi k t\} \tag{3.64}$$

$$\tilde{F}_{21} \sim \exp\{-k + 2\xi k t\} \tag{3.65}$$

$$\tilde{F}_{22} \sim \exp\{-2(\xi + k) + 2\xi k t\} \tag{3.66}$$

Eqs. (3.63)-(3.66) *seemingly* demonstrate the instability of both boundaries, i.e. the exponential growth in time with the increment $2\xi k$ (see also the remark at the beginning of Section II). Indeed, the perturbations of the boundaries are governed by a ***linear system***; a linear system could be either stable or unstable ***as a whole***; so both boundaries with necessity should be formally either stable or unstable if their interaction is taken into account. However, comparing (3.63)-(3.66) it is easily seen that the ratio of the perturbation amplitudes at the oxidizing side to that at the reducing side, $\tilde{F}_{21}/\tilde{F}_{11}$ and $\tilde{F}_{22}/\tilde{F}_{11}$ decreases exponentially with the wave number $k$. This means that the boundary at the oxidizing side is stable for perturbations with wavelengths smaller than the width of the layer, i.e. it is practically morphologically stable, which is in complete agreement with both theoretical consideration and experimental observations in [9]. We would like to point out that this result for the linear stability of the boundaries (for zero surface tension) is ***exact***. The diffusional interaction of the boundaries is taken into account; till now we have not used any additional assumptions. The study of the morphological stability of boundaries for the stationary solution is thereby reduced to exploration of the singular points (in the complex plane) of the corresponding integrands.

## IV. QUASI-STATIONARY APPROXIMATION

While being exact, the approach used in the previous Section is quite complicated. So it appears reasonable to check the simpler but approximate way of solution of the same problem, compare the results, and then use the simpler approach to solve the essentially more complicated problem for the two-layer system. The approximate approach is based on the physical fact that the deviations from the stoichiometry at the boundaries are quite small, $\delta_1 < \delta_2 \ll 1$; then $\xi$ is a small parameter too

$$\xi = \frac{1}{2} \ln \frac{1-\delta_1}{1-\delta_2} \ll 1 \tag{4.1}$$

This means that the characteristic time for the development of the instability $\frac{1}{2\xi k}$, see (3.57) is large as compared to the characteristic time of the diffusional relaxation inside the layer (which we have taken as a time scale, see Section 2). Then we can use the quasi-stationary approximation for (3.21), that is, drop the time derivative:

$$2\xi \frac{\partial w}{\partial z} + \frac{\partial^2 w}{\partial z^2} - \left(k^2 + \eta\right) w = 0 \tag{4.2}$$

In this approximation the time evolution enters via the boundary conditions (3.24)-(3.25), see below. We still consider the zero surface tension case, $\Gamma_1 = \Gamma_2 = 0$; the boundary conditions (3.22)-(3.23) become

$$w\big|_{z=0} + 2\xi\left(1-\delta_1\right)\gamma_1\left(k,t\right) = 0 \tag{4.3}$$

$$w\big|_{z=1} + 2\xi\left(1-\delta_2\right)\gamma_2\left(k,t\right) = 0 \tag{4.4}$$

The solution of the quasistationary problem is given in Appendix 4; here we only outline the procedure and give the result. First, the solution of (4.2) satisfying the boundary conditions (4.3)-(4.4) is obtained. Then it is substituted into the boundary conditions (3.24)-(3.25); in this way we obtain the system of two ordinary linear equations ((10.4)-(10.5) in Appendix 4) governing the time evolution of the boundary perturbations $\gamma_1(k,t)$ and $\gamma_2(k,t)$. Solving this system in a standard way, considering $k$ larger than one (see (10.13)-(10.14)), and returning to the Fourier modes of the boundaries' perturbations $\bar{\varphi}_1(k,t)$ and $\bar{\varphi}_2(k,t)$, yields the following expressions for their time evolution

$$\bar{\varphi}_1(k,t) = A_1 \exp(2\xi kt), \quad A_1 \simeq \bar{\varphi}_1(0,k) - \exp(-2\xi - k)\bar{\varphi}_2(0,k) \quad (4.5)$$

$$\begin{aligned}\bar{\varphi}_2(k,t) &= B_1 \exp(2\xi kt), \\ B_1 &\simeq \exp(-k)\bar{\varphi}_1(k,0) - \exp(-2(\xi+k))\bar{\varphi}_2(k,0),\end{aligned} \quad (4.6)$$

which coincides with the exact results of the previous section, Eqs.(3.61)-(3.66). I.e., the approximate analysis of morphological stability based on the quasi-stationary solution yields essentially the same results as the exact approach (for the perturbations with the wave length smaller then the width of the layer). It is worth mentioning that for a single boundary the quasi-stationary approach is exactly equivalent to the method used in [9]. Now it is evident that for a multilayer system the ratio of characteristic times will not change qualitatively, so application of the much simpler approximate method is again justified. In the next Section we study the coupled morphological stability of three boundaries using the quasistationary approach.

## V. STABILITY ANALYSIS FOR TWO OXIDE LAYERS

For two oxide layers, that is three interphase boundaries (see Fig. 2), the governing equations and boundary conditions for the first order perturbations are, see (2.8)–(2.23):

$$\frac{1}{D_\alpha}\frac{\partial u}{\partial \tau} = \frac{V}{D_\alpha}\frac{\partial u}{\partial Z} + \frac{\partial^2 u}{\partial X^2} + \frac{\partial^2 u}{\partial Z^2}, \tag{5.1}$$

$$\frac{1}{D_\beta}\frac{\partial v}{\partial \tau} = \frac{V}{D_\beta}\frac{\partial v}{\partial Z} + \frac{\partial^2 v}{\partial X^2} + \frac{\partial^2 v}{\partial Z^2}, \tag{5.2}$$

$$u\Big|_{Z=0} + \frac{V}{D_\alpha}\frac{1-\delta_1^\alpha}{\omega_\alpha}\Phi_1(X,\tau) + \frac{\delta_1^\alpha}{\omega_\alpha}\tilde{\Gamma}_1\frac{\partial^2 \Phi_1}{\partial X^2} = 0, \tag{5.3}$$

$$u\Big|_{Z=L_\alpha} + \frac{V}{D_\alpha}\frac{1-\delta_2^\alpha}{\omega_\alpha}\Phi_2(X,\tau) - \frac{\delta_2^\alpha}{\omega_\alpha}\tilde{\Gamma}_2\frac{\partial^2 \Phi_2}{\partial X^2} = 0, \tag{5.4}$$

$$v\Big|_{Z=L_\alpha} + \frac{V}{D_\beta}\frac{3-\delta_2^\beta}{\omega_\beta}\Phi_2(X,\tau) + \frac{\delta_2^\beta}{\omega_\beta}\tilde{\Gamma}_2\frac{\partial^2 \Phi_2}{\partial X^2} = 0, \tag{5.5}$$

$$v\Big|_{Z=L_\alpha+L_\beta} + \frac{V}{D_\beta}\frac{3-\delta_3^\beta}{\omega_\beta}\Phi_3(X,\tau) - \frac{\delta_3^\beta}{\omega_\beta}\tilde{\Gamma}_3\frac{\partial^2 \Phi_3}{\partial X^2} = 0, \tag{5.6}$$

$$\frac{1}{D_\alpha}\frac{\partial \Phi_1}{\partial \tau} = \frac{\omega_\alpha}{1-\delta_1^\alpha}\frac{\partial u}{\partial Z}\Big|_{Z=0} - \left(\frac{V}{D_\alpha}\right)^2 \Phi_1(X,\tau) \tag{5.7}$$

$$\frac{\partial \Phi_2}{\partial \tau} = \left[\frac{3-\delta_2^\beta}{\omega_\beta} - \frac{1-\delta_2^\alpha}{\omega_\alpha}\right]^{-1}\left[D_\beta\left(\frac{\partial v}{\partial Z}\Big|_{Z=L_\alpha} - \frac{3-\delta_2^\beta}{\omega_\beta}\left(\frac{V}{D_\beta}\right)^2\Phi_2(X,\tau)\right) - \right.$$
$$\left. -D_\alpha\left(\frac{\partial u}{\partial Z}\Big|_{Z=L_\alpha} - \frac{1-\delta_2^\alpha}{\omega_\alpha}\left(\frac{V}{D_\alpha}\right)^2\Phi_2(X,\tau)\right)\right] \tag{5.8}$$

$$\frac{1}{D_\beta}\frac{\partial \Phi_3}{\partial \tau} = \frac{\omega_\beta}{3-\delta_3^\beta}\frac{\partial v}{\partial Z}\bigg|_{Z=L_\alpha+L_\beta} - \left(\frac{V}{D_\beta}\right)^2 \Phi_3(X,\tau) \tag{5.9}$$

Taking the stationary width $L_\alpha$ of the $\alpha$-phase layer as the length scale and, correspondingly rescaling all other lengths $Z/L_\alpha = z$, $X/L_\alpha = x$, $\Phi_i/L_\alpha = \varphi_i$, $\tilde{\Gamma}_i/L_\alpha = \Gamma_i$, and time $\tau/\left(\frac{L_\alpha^2}{D_\alpha}\right) = t$, measuring $u, v$ in molar fractions $\omega_\alpha u = \tilde{u}$, $\omega_\beta v = \tilde{v}$, and defining $l = L_\beta/L_\alpha$, we are led to the dimensionless system of equations for the perturbations of concentration fields

$$\frac{\partial \tilde{u}}{\partial t} = 2\xi \frac{\partial \tilde{u}}{\partial z} + \frac{\partial^2 \tilde{u}}{\partial x^2} + \frac{\partial^2 \tilde{u}}{\partial z^2}; \quad 0 < z < 1, \tag{5.10}$$

$$\theta\frac{\partial \tilde{v}}{\partial t} = 2\xi\theta \frac{\partial \tilde{v}}{\partial z} + \frac{\partial^2 \tilde{v}}{\partial x^2} + \frac{\partial^2 \tilde{v}}{\partial z^2}; \quad 1 < z < l \tag{5.11}$$

where

$$\xi = \frac{VL_\alpha}{2D_\alpha} = \frac{1}{2}\ln\frac{1-\delta_1^\alpha}{1-\delta_2^\alpha}; \quad \theta = \frac{D_\alpha}{D_\beta}. \tag{5.12}$$

The boundary conditions (5.3)-(5.9) become

$$\tilde{u}\bigg|_{z=0} + 2\xi(1-\delta_1^\alpha)\varphi_1(x,t) + \delta_1^\alpha \Gamma_1 \frac{\partial^2 \varphi_1}{\partial x^2} = 0, \tag{5.13}$$

$$\tilde{u}\bigg|_{z=1} + 2\xi(1-\delta_2^\alpha)\varphi_2(x,t) - \delta_2^\alpha \Gamma_2 \frac{\partial^2 \varphi_2}{\partial x^2} = 0, \tag{5.14}$$

$$\tilde{v}\bigg|_{z=1} + 2\xi\theta(3-\delta_2^\beta)\varphi_2(x,t) + \delta_2^\beta \Gamma_2 \frac{\partial^2 \varphi_2}{\partial x^2} = 0, \tag{5.15}$$

$$\tilde{v}\Big|_{z=1+l} + 2\xi\theta(3-\delta_3^\beta)\varphi_3(x,t) - \delta_3^\beta \Gamma_3 \frac{\partial^2 \varphi_3}{\partial x^2} = 0, \tag{5.16}$$

$$\frac{\partial \varphi_1}{\partial t} = \frac{1}{1-\delta_1^\alpha} \frac{\partial \tilde{u}}{\partial z}\Big|_{z=0} - 4\xi^2 \varphi_1(x,t), \tag{5.17}$$

$$(\rho-1)\frac{\partial \varphi_2}{\partial t} = \frac{\rho}{\theta(3-\delta_2^\beta)} \frac{\partial \tilde{v}}{\partial z}\Big|_{z=1} - \frac{1}{1-\delta_2^\alpha} \frac{\partial \tilde{u}}{\partial z}\Big|_{z=1} + (1-\rho\theta)4\xi^2 \varphi_2(x,t), \tag{5.18}$$

where

$$\rho = \frac{(3-\delta_2^\beta)\omega_\alpha}{\omega_\beta(1-\delta_2^\alpha)} \tag{5.19}$$

is the ratio of the metal atoms equilibrium concentrations (per unit volume) for two oxides at their common boundary, and

$$\theta \frac{\partial \varphi_3}{\partial t} = \frac{1}{3-\delta_3^\beta} \frac{\partial \tilde{v}}{\partial z}\Big|_{z=1+l} - 4\theta^2 \xi^2 \varphi_3(x,t), \tag{5.20}$$

Introducing the Fourier transforms,

$$\varphi_j(x,t) = \frac{1}{\sqrt{2\pi}} \int_{-\infty}^{\infty} dk \exp(ikx) \overline{\varphi}_j(k,t), \; j=1,2,3 \tag{5.21}$$

$$\tilde{u}(x,z,t) = \frac{1}{\sqrt{2\pi}} \int_{-\infty}^{\infty} dk \exp(ikx) \overline{u}(k,z,t), \tag{5.22}$$

$$\tilde{v}(x,z,t) = \frac{1}{\sqrt{2\pi}} \int_{-\infty}^{\infty} dk \exp(ikx) \overline{v}(k,z,t), \tag{5.23}$$

We obtain from (5.10)-(5.11), (5.13)-(5.18), and (5.20):

$$\frac{\partial \overline{u}}{\partial t} = 2\xi \frac{\partial \overline{u}}{\partial z} + \frac{\partial^2 \overline{u}}{\partial z^2} - k^2 \overline{u}, \tag{5.24}$$

$$\theta \frac{\partial \overline{v}}{\partial t} = 2\xi\theta \frac{\partial \overline{v}}{\partial z} + \frac{\partial^2 \overline{v}}{\partial x^2} - k^2 \overline{v} \tag{5.25}$$

$$\overline{u}|_{z=0} + 2\xi(1-\delta_1^\alpha)\overline{\varphi}_1 - \delta_1^\alpha \Gamma_1 k^2 \overline{\varphi}_1 = 0, \tag{5.26}$$

$$\overline{u}|_{z=1} + 2\xi(1-\delta_2^\alpha)\overline{\varphi}_2 + \delta_2^\alpha \Gamma_2 k^2 \overline{\varphi}_2 = 0, \tag{5.27}$$

$$\overline{v}|_{z=1} + 2\xi\theta(3-\delta_2^\beta)\overline{\varphi}_2 - \delta_2^\beta \Gamma_2 k^2 \overline{\varphi}_2 = 0, \tag{5.28}$$

$$\overline{v}|_{z=1+l} + 2\xi\theta(3-\delta_3^\beta)\overline{\varphi}_3 + \delta_3^\beta \Gamma_3 k^2 \overline{\varphi}_3 = 0, \tag{5.29}$$

$$\frac{\partial \overline{\varphi}_1}{\partial t} = \frac{1}{1-\delta_1^\alpha} \frac{\partial \overline{u}}{\partial z}\bigg|_{z=0} - 4\xi^2 \overline{\varphi}_1(k,t), \tag{5.30}$$

$$(\rho-1)\frac{\partial \overline{\varphi}_2}{\partial t} = \frac{\rho}{\theta(3-\delta_2^\beta)} \frac{\partial \overline{v}}{\partial z}\bigg|_{z=1} - \frac{1}{1-\delta_2^\alpha} \frac{\partial \overline{u}}{\partial z}\bigg|_{z=1} + (1-\rho\theta)4\xi^2 \overline{\varphi}_2(k,t), \tag{5.31}$$

$$\theta \frac{\partial \overline{\varphi}_3}{\partial t} = \frac{1}{3-\delta_3^\beta} \frac{\partial \overline{v}}{\partial z}\bigg|_{z=1+l} - 4\theta^2 \xi^2 \overline{\varphi}_3(k,t), \tag{5.32}$$

Following the approach in Section 3, we again introduce a new variable that is we add a fictitious "dissipation", which may be adjusted to compensate the possible instability:

$$w^\alpha = \overline{u}\exp(-\eta t), \tag{5.33}$$

$$w^\beta = \overline{v}\exp(-\eta t), \tag{5.34}$$

$$\gamma_i = \overline{\varphi}_i \exp(-\eta t), i=1,2,3, \tag{5.35}$$

where the constant $\eta > 0$ is undetermined; the upper index $\alpha$ refers to the AO layer ($\alpha$-phase), and the index $\beta$ refers to the $A_3O_4$ layer ($\beta$-phase). In terms of these new variables (5.24)-(5.25) become:

$$\frac{\partial w^{\alpha}}{\partial t} = 2\xi \frac{\partial w^{\alpha}}{\partial z} + \frac{\partial^2 w^{\alpha}}{\partial z^2} - \left(k^2 + \eta\right) w^{\alpha}, \tag{5.36}$$

$$\theta \frac{\partial w^{\beta}}{\partial t} = 2\xi\theta \frac{\partial w^{\beta}}{\partial z} + \frac{\partial^2 w^{\beta}}{\partial z^2} - \left(k^2 + \eta\right) w^{\beta}. \tag{5.37}$$

The boundary conditions (5.26)-(5.29) do not change their form:

$$w^{\alpha}\big|_{z=0} + \left[2\xi(1-\delta_1^{\alpha}) - \delta_1^{\alpha}\Gamma_1 k^2\right]\gamma_1 = 0, \tag{5.38}$$

$$w^{\alpha}\big|_{z=1} + \left[2\xi(1-\delta_2^{\alpha}) + \delta_2^{\alpha}\Gamma_2 k^2\right]\gamma_2 = 0, \tag{5.39}$$

$$w^{\beta}\big|_{z=1} + \left[2\xi\theta(3-\delta_2^{\beta}) - \delta_2^{\beta}\Gamma_2 k^2\right]\gamma_2 = 0, \tag{5.40}$$

$$w^{\beta}\big|_{z=1+l} + \left[2\xi\theta(3-\delta_3^{\beta}) + \delta_3^{\beta}\Gamma_3 k^2\right]\gamma_3 = 0, \tag{5.41}$$

and the boundary conditions (5.30)-(5.32) become

$$\frac{\partial \gamma_1}{\partial t} = \frac{1}{1-\delta_1^{\alpha}} \frac{\partial w^{\alpha}}{\partial z}\bigg|_{z=0} - \left[4\xi^2 + \eta\right]\gamma_1(k,t), \tag{5.42}$$

$$(\rho-1)\frac{\partial \gamma_2}{\partial t} = \frac{\rho}{\theta(3-\delta_2^{\beta})} \frac{\partial w^{\beta}}{\partial z}\bigg|_{z=1} - \frac{1}{1-\delta_2^{\alpha}} \frac{\partial w^{\alpha}}{\partial z}\bigg|_{z=1} + \\ + \left[(1-\rho)\eta + (1-\rho\theta)4\xi^2\right]\gamma_2(k,t) \tag{5.43}$$

$$\theta \frac{\partial \gamma_3}{\partial t} = \frac{1}{3-\delta_3^{\beta}} \frac{\partial w^{\beta}}{\partial z}\bigg|_{z=1+l} - \left[4\theta^2\xi^2 + \theta\eta\right]\gamma_3(k,t), \tag{5.44}$$

To make our further considerations most transparent we take again $\Gamma_1 = \Gamma_2 = \Gamma_3 = 0$,

and, additionally $\theta = 1$. That is we presume equal diffusion coefficients for both oxide phases as this simplifies all formulae essentially, in analogy to the 'Stationary symmetric model' of Langer [14]. The influence of the different diffusion coefficients and non-zero surface tension will be studied elsewhere. In the quasistationary approximation instead of (5.36)-(5.37) we consider

$$\frac{\partial^2 w^\alpha}{\partial z^2} + 2\xi \frac{\partial w^\alpha}{\partial z} - \left(k^2 + \eta\right) w^\alpha = 0 \tag{5.45}$$

$$\frac{\partial^2 w^\beta}{\partial z^2} + 2\xi \frac{\partial w^\beta}{\partial z} - \left(k^2 + \eta\right) w^\beta = 0 \tag{5.46}$$

with the boundary conditions, see (5.38)-(5.41),

$$w^\alpha \bigg|_{z=0} = -2\xi(1 - \delta_1^\alpha)\gamma_1 \tag{5.47}$$

$$w^\alpha \bigg|_{z=1} = -2\xi(1 - \delta_2^\alpha)\gamma_2 \tag{5.48}$$

$$w^\beta \bigg|_{z=1} = -2\xi(3 - \delta_2^\beta)\gamma_2 \tag{5.49}$$

$$w^\beta \bigg|_{z=1+l} = -2\xi(3 - \delta_3^\beta)\gamma_3 \tag{5.50}$$

Here we only outline the most essential steps of the quite tedious solution procedure, moving all the details to the Appendix 5. First, the solutions of (5.45)-(5.46) satisfying the boundary conditions (5.47)-(5.50) are obtained. Then the values of the derivatives $\dfrac{\partial w^\alpha}{\partial z}$ and $\dfrac{\partial w^\beta}{\partial z}$ at the corresponding boundaries are calculated and substituted into the boundary conditions (5.42)-(5.44); in this way we obtain the system of three ordinary linear equations ((11.7)-(11.9) in Appendix 5) governing the time evolution of the renormalized boundary perturbations $\gamma_1(k,t), \gamma_2(k,t)$ and $\gamma_3(k,t)$. Solution of this system is obtained in the same way as for the case of two boundaries: substitution of $\gamma_1 = A\exp(\sigma t)$, $\gamma_2 = B\exp(\sigma t)$, and $\gamma_3 = C\exp(\sigma t)$ yields a linear homogeneous algebraic system for $A, B$, and $C$. For solutions of this system to exist the determinant (11.21) of this system should equal zero. After some algebra this yields the following equation for $\sigma$

$$\det(G_{ij}) = \left[\left(2\xi^2 + \eta + \sigma\right)^2 - \left(2\xi\zeta\right)^2\right] \times \\ \times \left[2\xi\zeta\,cth\zeta l + 2\rho\xi\zeta\,cth\zeta + (1-\rho)\left(2\xi^2 + \eta + \sigma\right)\right] = 0 \tag{5.51}$$

The very fact that the determinant (11.21) (with the elements given by (11.14)-(11.20)) is **exactly** reduced to the short expression (5.51), which is additionally decomposed in two parts, is remarkable! This is a direct consequence of the special feature of the stationary solutions (10.1), and (11.1)-(11.2): that the width of the layer and velocity are expressed via boundary conditions, and vice versa.

This equation has three roots: two roots are given by

$$\left(2\xi^2 + \eta + \sigma\right)^2 - \left(2\xi\zeta\right)^2 = 0 \tag{5.52}$$

and the third by

$$2\xi\zeta\,cth\zeta l + 2\rho\xi\zeta\,cth\zeta + (1-\rho)\left(2\xi^2 + \eta + \sigma\right) = 0 \tag{5.53}$$

Now, $\rho$ is the ratio of the metal atoms equilibrium concentration (per unit volume) for the higher oxide to that for the lower oxide at their common boundary, see (5.19); that is in most cases, e.g. $AO$ and $A_3O_4$ (A = Co, Ni, Fe….) $\rho < 1$. This means that the root of (5.53) is always negative, so only roots of (5.52) are of interest from the point of the possible instability:

$$\sigma_{1,2} = -\left(2\xi^2 + \eta\right) \pm 2\xi\zeta \tag{5.54}$$

The margin of stability corresponds to zero value of the largest root, $\sigma_1 = 0$. In its turn this means $\eta = 2\xi k$. Again, as in the previous Section, in terms of the Fourier modes of the boundaries' perturbations $\bar{\varphi}_i, i = 1,2$ this means the increment $\eta = 2\xi k$ for the $k$-th mode. Correspondingly, to compare the amplitudes at the onset of instability we need only $A_1, B_1$ and $C_1$. Exact (quite complicated) expressions for these amplitudes are given in Appendix 5. For the sake of simplicity, we discuss here only the case when $l = L_\beta/L_\alpha$ is not a small parameter, which means that the stationary widths of the layers are comparable. Then for $k$ larger than one, i.e. for the perturbations with the wave length smaller than the widths of the layers, these expressions simplify drastically, see (11.43)-(11.45), and in terms the Fourier modes of the boundaries' perturbations $\bar{\varphi}_1(k,t)$, $\bar{\varphi}_2(k,t)$ and $\bar{\varphi}_3(k,t)$ we obtain:

$$\begin{aligned}&\bar{\varphi}_1(k,t) = A_1 \exp(2\xi k); \\ &A_1 \simeq \bar{\varphi}_1(k,0) - (1-\rho)\exp(-2\xi - k)\bar{\varphi}_2(k,0) - \\ &-\rho \exp(-(2+l)\xi - k)\bar{\varphi}_1(k,0)\end{aligned} \tag{5.55}$$

$$\begin{aligned}&\bar{\varphi}_2(k,t) = B_1 \exp(2\xi k); \\ &B_1 \simeq \exp(-k)\bar{\varphi}_1(k,0) - (1-\rho)\exp(-2(\xi+k))\bar{\varphi}_2(k,0) - \\ &-\rho \exp(-(2+l)\xi - 2k)\bar{\varphi}_3(k,0)\end{aligned} \tag{5.56}$$

$$\begin{aligned}&\bar{\varphi}_3(k,t) = C_1 \exp(2\xi k); \\ &C_1 \simeq \exp(-k(1+l))\bar{\varphi}_1(k,0) - (1-\rho)\exp(-2\xi - k(2+l))\bar{\varphi}_2(k,0) - \\ &-\rho \exp(-(2+l)\xi - k(2+l))\bar{\varphi}_3(k,0)\end{aligned} \tag{5.57}$$

It's worth mentioning that for $\rho = 0$, i.e. only for a single oxide layer, $A_1$ in (5.55) and $B_1$

in (5.56) are reduced to

$$A_1 \simeq \gamma_1(0,k) - \exp(-2\xi - k)\gamma_2(0,k) \tag{5.58}$$

$$B_1 \simeq \exp(-k)\gamma_1(0,k) - \exp(-2(\xi+k))\gamma_2(0,k) \tag{5.59}$$

i.e. to (4.5)-(4.6).

It is evident from (5.55)-(5.57) that for the double oxide layer, e.g. $AO/A_3O_4$, the surface on the reducing side is again unstable. The stability of the intermediate boundary is practically the same (up to coefficient of order unity) as of the oxidizing surface for the single-oxide case. On the other hand the surface of the higher oxide on oxidizing side is even much more morphologically stable.

## VI. SUMMARY AND CONCLUSIONS

In this paper, we have studied the coupled morphological stability of multiple phase boundaries for oxides that are exposed to an oxygen potential gradient. For a single oxide layer this problem was considered in [9], both experimentally and theoretically. It was shown that while the oxidizing boundary is morphologically stable, the reducing boundary becomes unstable. In the present work the problem of [9] is generalized in two ways: first, in exploring the morphological stability of two solid/gas interfaces their diffusional interaction is taken into account; second we consider two oxide layers with two solid/das interafecs and one solid/solid interface. To explore the stability of diffusionally interacting boundaries the method developed in [10, 11] is applied. Based on integral transformation of a special kind this method reveals the evolution of the boundaries' perturbations *without solving the diffusional problem inside the layer*. The study of the morphological stability of boundaries for the stationary solution is thereby reduced to exploration of the singular points (in the complex plane) of the corresponding integrands.

As it was mentioned above, from the formal point the result of [9] may look paradoxical: if the interaction of the boundaries is taken into account the perturbations of the boundaries are governed by a coupled *linear* system; a linear system could be either stable, or unstable *as a whole*; so formally both boundaries with necessity should be either stable or unstable. However, comparing (3.57)-(3.60) it is easily seen that the ratio of the perturbation amplitudes at the oxidizing side to that at the reducing side decreases exponentially with the wave number $k$. This means that the boundary at the oxidizing side is practically morphologically stable indeed, which is in complete agreement with both theoretical consideration and experimental observations in [9]. To visualize the mutual influence of the boundaries' perturbations it is practical to plot the $\ln\left|\tilde{F}_{ij} / \tilde{F}_{11}\right|$ (see (3.57) - (3.60)) against the wave number $k$ (see Fig 5).

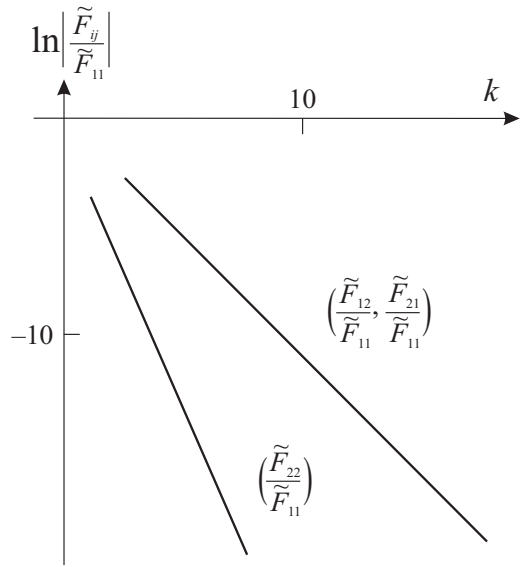

FIG. 5. Plot of $\ln\left|\tilde{F}_{ij} / \tilde{F}_{11}\right|$ against the wave number $k$ for the case of a single layer.

Because $\xi \ll 1$, the "cross-influences" of the perturbations of reducing and oxidizing boundaries, $\tilde{F}_{12}/\tilde{F}_{11}$ and , $\tilde{F}_{21}/\tilde{F}_{11}$ , exhibit identical asymptotical behavior in $-k$, while the "self-influence" of the oxidizing boundary, $\tilde{F}_{22}/\tilde{F}_{11}$ decreases as $-2k$. It is remarkable that these results for the linear stability of the boundaries of a single layer are not dependent on the width of the layer and (for zero surface tension) are ***exact***. On the other hand, while being exact, the above approach is quite complicated. So our strategy was to check the simpler but approximate way of solution of the same (single-layer) problem, compare the results, and then use the simpler approach to solve the essentially more complicated problem for the two-layer system. The approximate approach is based on the physical fact that the deviations from the stoichiometry at the boundaries are quite small and, consequently, $\xi$ (see eq.(4.1)) is a small parameter. This means that the characteristic time for the development of the instability $(2\xi k)^{-1}$ is large as compared to the characteristic time of the diffusional relaxation inside the layer (which we have taken as a time scale, see Section 2). Therefore the quasi-stationary approximation is justified; the approximate analysis of morphological stability based on the quasi-stationary solution yields essentially the same results as the exact approach. It is worth mentioning that for a single boundary the quasi-stationary approach is exactly equivalent to the method used in [9]. Now it is evident that for a two-layer system the ratio of characteristic times will not change qualitatively (if the diffusion coefficients in the layers are not too different), so application of the much simpler approximate method is again justified. Here we have studied the coupled morphological stability of three boundaries for the case of equal diffusion coefficients (symmetrical model) in both layers. Then the quasistationary approach reveals that the surface at the reducing side is again unstable. The stability of the intermediate boundary is practically the same (up to coefficient of order unity) as of the oxidizing surface for the single-oxide case. On the other hand the surface of the higher oxide on oxidizing side is even much more morphologically stable. So to obtain a morphologically stable oxide layer it may be expedient to grow it on the base of lower oxide. Again, to visualize the mutual influence of the boundaries' perturbations it is practical to plot the $\ln\left|\tilde{F}_{ij}/\tilde{F}_{11}\right|$ against the wave number $k$; here the $\tilde{F}_{ij}$, $i,j = 1,2,3$ are introduced for three boundaries exactly in the same way, as they were introduced in (3.61)-(3.66) for two. However, there is now a dependence on the ratio of the layer widths $l = L_\beta/L_\alpha$: for $l < 1$ see Fig 6, and for $l > 1$, see Fig 7.

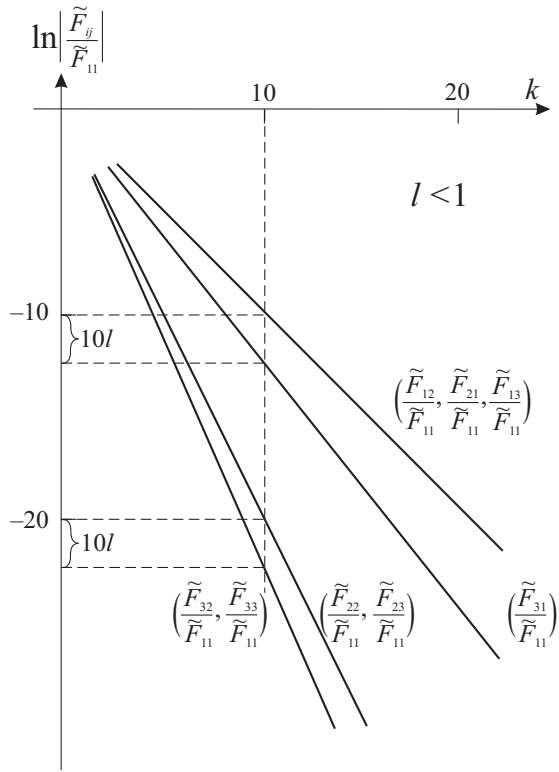
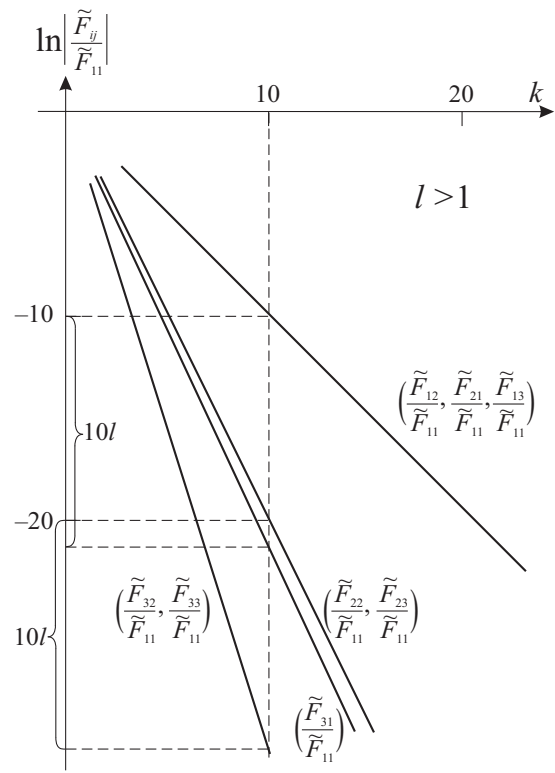

FIG. 6. The $k$-dependence of $\ln\left|\tilde{F}_{ij}/\tilde{F}_{11}\right|$ for the case $l < 1$.

FIG. 7. The $k$-dependence of $\ln\left|\tilde{F}_{ij}/\tilde{F}_{11}\right|$ for the case $l > 1$.

Thus, for $l < 1$ both the self-influence of the solid/solid boundary and the influence of the oxidizing boundary on it decrease faster with the wave number than the cross-influence of the reducing surface on the oxidizing surface. The situation is inversed when $l > 1$, i.e. when the width $L_\beta$ is larger than the width $L_\alpha$.

## ACKNOWLEDGEMENT

We acknowledge the help of Dr. L. Davydov in preparing the figures.

## APPENDIX 1

If we look for the stationary ("zero order") solution, the equations (2.1) and (2.4)-(2.7), are reduced to [9]

$$-V \frac{dC_s}{dZ} - D \frac{d^2C_s}{dZ^2} = 0, \qquad (7.1)$$

$$C_s \Big|_{Z=0} = C_1 = \frac{\delta_1}{\omega}, \quad C_s \Big|_{Z=L} = C_2 = \frac{\delta_2}{\omega}, \qquad (7.2)$$

$$V = \frac{\omega D}{1-\delta_1} \frac{dC_s}{dZ} \Big|_0 = \qquad (7.3)$$

$$= \frac{\omega D}{1-\delta_2} \frac{dC_s}{dZ} \Big|_L, \qquad (7.4)$$

The solution of (7.1), satisfying the boundary conditions (7.2), is [9]:

$$C_s = \frac{C_2 - C_1 \exp(-VL/D) - (C_2 - C_1)\exp(-VZ/D)}{1 - \exp(-VL/D)} \qquad (7.5)$$

Now we have two more equations, (7.3) and (7.4), to determine $V$ and the stationary width $L$. Substitution of (7.5) for $C$ into (7.3)-(7.4) yields:

$$1 - \delta_1 = \omega \frac{C_2 - C_1}{1 - \exp(-VL/D)}, \tag{7.6}$$

$$1 - \delta_2 = \omega \frac{C_2 - C_1}{1 - \exp(-VL/D)} \exp(-VL/D). \tag{7.7}$$

It follows from (7.6)-(7.7):

$$\exp\left(\frac{VL}{D}\right) = \frac{1 - \delta_1}{1 - \delta_2}. \tag{7.8}$$

The interesting feature of the present solution is that only the product of $V$ and $L$ is determined by (7.8), but not each of these quantities separately (this reminds on the well-known "velocity selection controversy" [12].) However, in the present case we have an additional physically motivated condition: the total amount of $A$ atoms is conserved, that is [9]

$$\int_0^{L_0} dZ \frac{1 - \delta_0}{\omega} = \int_0^L dZ \left(\frac{1}{\omega} - C_s(Z)\right). \tag{7.9}$$

where $L_0$ is the initial oxide layer thickness. Substitution of $C_s(Z)$, see (7.5), into the latter equation yields finally [9]:

$$L = L_0 \frac{\frac{1}{\omega} - C_0}{C_2 - C_1} \ln \frac{1 - C_1 \omega}{1 - C_2 \omega} \tag{7.10}$$

and

$$V = \frac{D}{L_0} \frac{C_2 - C_1}{\frac{1}{\omega} - C_0} = \frac{D}{L_0} \frac{\delta_2 - \delta_1}{1 - \delta_0}. \tag{7.11}$$

One can easily check that in the limit $(C_2 - C_1) \to 0$ (and, correspondingly, $V \to 0$ ) the stationary width $L$ approaches the initial width $L_0$.

For the subsequent stability analysis we will need to know the values of the derivatives $\dfrac{\partial C_s}{\partial Z}$ and $\dfrac{\partial^2 C_s}{\partial Z^2}$ at the boundaries $Z = 0$ and $Z = L$,

$$\left.\frac{\partial C_s}{\partial Z}\right|_{Z=0} = \frac{V}{D}\frac{1-\delta_1}{\omega}, \quad \left.\frac{\partial^2 C_s}{\partial Z^2}\right|_{Z=0} = -\left(\frac{V}{D}\right)^2 \frac{1-\delta_1}{\omega}, \qquad (7.12)$$

$$\left.\frac{\partial C_s}{\partial Z}\right|_{Z=L} = \frac{V}{D}\frac{1-\delta_2}{\omega}, \quad \left.\frac{\partial^2 C_s}{\partial Z^2}\right|_{Z=L} = -\left(\frac{V}{D}\right)^2 \frac{1-\delta_2}{\omega}. \qquad (7.13)$$

## Appendix 2

If we look for the stationary ("zero order") solution which is only $Z$-dependent, quite analogous to Subsection 2.1, the equations (2.8)-(2.9), (2.15)-(2.18), and (2.21)-(2.23) are reduced to

$$V\frac{dC_s^\alpha}{dZ} + D_\alpha \frac{d^2 C_s^\alpha}{dZ^2} = 0, \ 0 < Z < L_\alpha \qquad (8.1)$$

$$V\frac{dC_s^\beta}{dZ} + D_\beta \frac{d^2 C_s^\beta}{dZ^2} = 0, \ L_\alpha < Z < L_\alpha + L_\beta \qquad (8.2)$$

$$\left.C_s^\alpha\right|_{Z=0} = C_{\alpha 1}; \left.C_s^\alpha\right|_{Z=L_\alpha} = C_{\alpha 2} \qquad (8.3)$$

$$\left.C_s^\beta\right|_{Z=L_\alpha} = C_{\beta 2}; \left.C_s^\beta\right|_{Z=L_\alpha+L_\beta} = C_{\beta 3} \qquad (8.4)$$

$$V = \frac{\omega_\alpha}{1-\delta_1^\alpha} D_\alpha \left.\frac{\partial C_s^\alpha}{\partial Z}\right|_{Z=0}, \qquad (8.5)$$

$$V = \left[\frac{3-\delta_2^\beta}{\omega_\beta} - \frac{1-\delta_2^\alpha}{\omega_\alpha}\right]^{-1} \left( D_\beta \left.\frac{\partial C_s^\beta}{\partial Z}\right|_{Z=L_\alpha} - D_\alpha \left.\frac{\partial C_s^\alpha}{\partial Z}\right|_{Z=L_\alpha} \right) \qquad (8.6)$$

$$V = \frac{\omega_\beta}{3-\delta_3^\beta} D_\beta \left.\frac{\partial C_s^\beta}{\partial Z}\right|_{Z=L_\alpha+L_\beta}, \qquad (8.7)$$

The solutions of the linear equations(8.1)-(8.2), satisfying boundary conditions(8.3)-(8.7) are obtained by straightforward generalization of (7.5). We will omit presenting here these lengthy expressions; for the exploration of stability we will need only the formulae for the widths of the layers (2.24)-(2.25), for the stationary velocity(2.26), and for the values of the derivatives $\dfrac{\partial C_s^\alpha}{\partial Z}$, $\dfrac{\partial C_s^\beta}{\partial Z}$ and $\dfrac{\partial^2 C_s^\alpha}{\partial Z^2}$, $\dfrac{\partial^2 C_s^\beta}{\partial Z^2}$ at corresponding boundaries:

$$\left.\frac{\partial C_s^\alpha}{\partial Z}\right|_{Z=0} = \frac{V}{D_\alpha}\frac{1-\delta_1^\alpha}{\omega_\alpha}; \quad \left.\frac{\partial^2 C_s^\alpha}{\partial Z^2}\right|_{Z=0} = -\left(\frac{V}{D_\alpha}\right)^2\frac{1-\delta_1^\alpha}{\omega_\alpha}, \tag{8.8}$$

$$\left.\frac{\partial C_s^\alpha}{\partial Z}\right|_{Z=L_\alpha} = \frac{V}{D_\alpha}\frac{1-\delta_2^\alpha}{\omega_\alpha}; \quad \left.\frac{\partial^2 C_s^\alpha}{\partial Z^2}\right|_{Z=L_\alpha} = -\left(\frac{V}{D_\alpha}\right)^2\frac{1-\delta_2^\alpha}{\omega_\alpha}, \tag{8.9}$$

$$\left.\frac{\partial C_s^\beta}{\partial Z}\right|_{Z=L_\alpha} = \frac{V}{D_\beta}\frac{3-\delta_2^\beta}{\omega_\beta}; \quad \left.\frac{\partial^2 C_s^\beta}{\partial Z^2}\right|_{Z=L_\alpha} = -\left(\frac{V}{D_\beta}\right)^2\frac{3-\delta_2^\beta}{\omega_\beta}, \tag{8.10}$$

$$\left.\frac{\partial C_s^\beta}{\partial Z}\right|_{Z=L_\alpha+L_\beta} = \frac{V}{D_\beta}\frac{3-\delta_3^\beta}{\omega_\beta}; \quad \left.\frac{\partial^2 C_s^\beta}{\partial Z^2}\right|_{Z=L_\alpha+L_\beta} = -\left(\frac{V}{D_\beta}\right)^2\frac{3-\delta_3^\beta}{\omega_\beta}. \tag{8.11}$$

**Appendix 3**

Here we give eight complex integrals $J_{lr}^{(m)}$, introduced in equation(3.55):

$$J_{11}^{(1)} = \int_{a-i\infty}^{a+i\infty} \frac{\left(\sqrt{y}+\xi+k\right)\exp(yt)}{\left(1-\exp\left(-2\sqrt{y}\right)\right)\left(y-(k+\xi)^2\right)\left(\sqrt{y}-\xi+k\right)}dy \tag{9.1}$$

$$J_{11}^{(2)} = \int_{a-i\infty}^{a+i\infty} \frac{\left(\sqrt{y}+k-\xi\right)\exp\left(yt-2\sqrt{y}\right)}{\left(1-\exp\left(-2\sqrt{y}\right)\right)\left(y-(k-\xi)^2\right)\left(\sqrt{y}+\xi+k\right)}dy \tag{9.2}$$

$$J_{12}^{(1)} = \int_{a-i\infty}^{a+i\infty} \frac{\left(\sqrt{y}+\xi+k\right)\exp\left(yt-\xi-\sqrt{y}\right)}{\left(1-\exp\left(-2\sqrt{y}\right)\right)\left[y-(k+\xi)^2\right]\left(\sqrt{y}-\xi+k\right)} dy \quad (9.3)$$

$$J_{12}^{(2)} = \int_{a-i\infty}^{a+i\infty} \frac{\left(\sqrt{y}+\xi+k\right)\exp\left(yt-\xi-\sqrt{y}\right)}{\left(1-\exp\left(-2\sqrt{y}\right)\right)\left[y-(k+\xi)^2\right]\left(\sqrt{y}-\xi+k\right)} dy \quad (9.4)$$

$$J_{21}^{(1)} = \int_{a-i\infty}^{a+i\infty} \frac{\left(\sqrt{y}+\xi+k\right)\exp\left(yt+\xi-\sqrt{y}\right)}{\left(1-\exp\left(-2\sqrt{y}\right)\right)\left[y-(k+\xi)^2\right]\left(\sqrt{y}-\xi+k\right)} dy \quad (9.5)$$

$$J_{21}^{(2)} = \int_{a-i\infty}^{a+i\infty} \frac{\left(\sqrt{y}+k-\xi\right)\exp\left(yt+\xi-\sqrt{y}\right)}{\left(1-\exp\left(-2\sqrt{y}\right)\right)\left[y-(k-\xi)^2\right]\left(\sqrt{y}+\xi+k\right)} dy \quad (9.6)$$

$$J_{22}^{(1)} = \int_{a-i\infty}^{a+i\infty} \frac{\left(\sqrt{y}+k-\xi\right)\exp\left(yt\right)}{\left(1-\exp\left(-2\sqrt{y}\right)\right)\left[y-(k-\xi)^2\right]\left(\sqrt{y}+\xi+k\right)} dy \quad (9.7)$$

$$J_{22}^{(2)} = \int_{a-i\infty}^{a+i\infty} \frac{\left(\sqrt{y}+\xi+k\right)\exp\left(yt-2\sqrt{y}\right)}{\left(1-\exp\left(-2\sqrt{y}\right)\right)\left[y-(k+\xi)^2\right]\left(\sqrt{y}-\xi+k\right)} dy \quad (9.8)$$

**Appendix 4**

The solution of (4.2) satisfying the boundary conditions (4.3)-(4.4) is easily obtained:

$$w = \frac{2\xi}{\exp(\lambda_1)-\exp(\lambda_2)}\{\left[(1-\delta_1)\exp(\lambda_2)\gamma_1-(1-\delta_2)\gamma_2\right]\exp(\lambda_1 z) + \\ +\left[(1-\delta_2)\gamma_2-(1-\delta_1)\exp(\lambda_1)\gamma_1\right]\exp(\lambda_2 z)\} \quad (10.1)$$

where $\lambda_1 = -\xi + \zeta$, $\lambda_2 = -\xi - \zeta$, and $\zeta = \sqrt{\xi^2 + k^2 + \eta}$. Below we need $\left.\dfrac{\partial w}{\partial z}\right|_{z=0}$ and $\left.\dfrac{\partial w}{\partial z}\right|_{z=1}$ only, that is the values of derivatives at the boundaries:

$$\frac{1}{1-\delta_1} \left.\frac{\partial w}{\partial z}\right|_{z=0} = 2\xi\left[\left(\xi + \zeta \coth \zeta\right)\gamma_1 - \frac{\zeta \exp(-\xi)}{\sinh \zeta}\gamma_2\right] \qquad (10.2)$$

$$\frac{1}{1-\delta_2} \left.\frac{\partial w}{\partial z}\right|_{z=1} = 2\xi\left[\left(\xi - \zeta \coth \zeta\right)\gamma_2 + \frac{\zeta \exp(\xi)}{\sinh \zeta}\gamma_1\right] \qquad (10.3)$$

Substituting these values into (3.24)-(3.25) we obtain a system of two equations for $\gamma_1(k,t)$, $\gamma_2(k,t)$:

$$\frac{\partial \gamma_1}{\partial t} = \left[2\xi\zeta \coth \zeta - 2\xi^2 - \eta\right]\gamma_1 - \frac{2\xi\zeta}{\sinh \zeta}\exp(-\xi)\gamma_2 \qquad (10.4)$$

$$\frac{\partial \gamma_2}{\partial t} = \frac{2\xi\zeta}{\sinh \zeta}\exp(\xi)\gamma_1 - \left[2\xi\zeta \coth \zeta + 2\xi^2 + \eta\right]\gamma_2 \qquad (10.5)$$

The solution of this system is obtained in a standard way: substitution of $\gamma_1 = A\exp(\sigma t)$, $\gamma_2 = B\exp(\sigma t)$ yields a linear homogeneous algebraic system for $A$, $B$

$$\left[2\xi\zeta \coth \zeta - 2\xi^2 - \eta - \sigma\right]A - \frac{2\xi\zeta}{\sinh \zeta}\exp(-\xi)B = 0 \qquad (10.6)$$

$$\frac{2\xi\zeta}{\sinh \zeta}\exp(\xi)A - \left[2\xi\zeta \coth \zeta + 2\xi^2 + \eta + \sigma\right]B = 0 \qquad (10.7)$$

For solutions of this system to exist the determinant of this system should equal zero, which yields after some algebra the quadratic equation for $\sigma$. The roots of the latter equation are

$$\sigma_{1,2} = -\left(2\xi^2 + \eta\right) \pm 2\xi\zeta \qquad (10.8)$$

The solution of the system (10.4)-(10.5) is

$$\gamma_1 = A_1 \exp(\sigma_1 t) + A_2 \exp(\sigma_2 t) \tag{10.9}$$

$$\gamma_2 = B_1 \exp(\sigma_1 t) + B_2 \exp(\sigma_2 t) \tag{10.10}$$

where $A_i, B_i$ are calculated in a standard way using initial values $\gamma_i(0, k)$. The margin of stability for $\gamma_i$ corresponds to zero value of the maximal (positive) root (10.8), which in its turn means $\eta = 2\xi k$. Again, in terms of the Fourier modes of the boundaries' perturbations $\bar{\varphi}_i, i = 1, 2$ this means the increment $\eta = 2\xi k$ for the $k$-th mode. Correspondingly, to compare the amplitudes at the onset of instability we need only $A_1$ and $B_1$

$$A_1 = \frac{1}{2}\left[1 + \coth(\xi + k)\right]\gamma_1(0, k) - \frac{\exp(-\xi)}{2\sinh(\xi + k)}\gamma_2(0, k) \tag{10.11}$$

$$B_1 = \frac{\exp(\xi)}{2\sinh(\xi + k)}\gamma_1(0, k) + \frac{1}{2}\left[1 - \coth(\xi + k)\right]\gamma_2(0, k) \tag{10.12}$$

Now, $k = 0$ means the shift of the layer as a whole; for $k \sim 1$ the ("transverse") scale of the perturbation is comparable to the width of the layer, which makes the use of quasistationary approximation problematical. On the other hand, for $\xi + k \gg 1$

$$A_1 \simeq \gamma_1(0, k) - \exp(-2\xi - k)\gamma_2(0, k) \tag{10.13}$$

$$B_1 \simeq \exp(-k)\gamma_1(0, k) - \exp(-2(\xi + k))\gamma_2(0, k) \tag{10.14}$$

which coincides with the exact results of the Section 3 ($k$ is in the argument of the exponent, so even when it equals 3 or 4 it is quite a reasonable approximation).

# APPENDIX 5

The solutions of (5.45)-(5.46) satisfying boundary conditions (5.47)-(5.50) are easily obtained

$$w^\alpha = \frac{2\xi}{\exp(\lambda_1) - \exp(\lambda_2)} \{[(1-\delta_1^\alpha)\exp(\lambda_2)\gamma_1 - (1-\delta_2^\alpha)\gamma_2]\exp(\lambda_1 z) + \\ + [(1-\delta_2^\alpha)\gamma_2 - (1-\delta_1^\alpha)\exp(\lambda_1)\gamma_1]\exp(\lambda_2 z)\} \quad (11.1)$$

$$w^\beta = \frac{2\xi}{\exp(\lambda_1 l) - \exp(\lambda_2 l)} \{[(3-\delta_2^\beta)\exp(\lambda_2 l)\gamma_2 - (3-\delta_3^\beta)\gamma_3]\exp(\lambda_1(z-1)) + \\ + [(3-\delta_3^\beta)\gamma_3 - (3-\delta_2^\beta)\exp(\lambda_1 l)\gamma_2]\exp(\lambda_2(z-1))\} \quad (11.2)$$

where $\lambda_1 = -\xi + \zeta$, $\lambda_2 = -\xi - \zeta$, and $\zeta = \left|\sqrt{\xi^2 + k^2 + \eta}\right|$. Below we need $\left.\frac{\partial w^\alpha}{\partial z}\right|_{z=0}$, $\left.\frac{\partial w^\alpha}{\partial z}\right|_{z=1}$, $\left.\frac{\partial w^\beta}{\partial z}\right|_{z=1}$, and $\left.\frac{\partial w^\beta}{\partial z}\right|_{z=1+l}$ only, that is the values of derivatives at the boundaries.

$$\frac{1}{1-\delta_1^\alpha}\left.\frac{\partial w^\alpha}{\partial z}\right|_{z=0} = 2\xi\left[(\xi + \zeta cth\zeta)\gamma_1 - \frac{\zeta\exp(-\xi)}{sh\zeta}\gamma_2\right] \quad (11.3)$$

$$\frac{1}{1-\delta_2^\alpha}\left.\frac{\partial w^\alpha}{\partial z}\right|_{z=1} = 2\xi\left[(\xi - \zeta cth\zeta)\gamma_2 + \frac{\zeta\exp(\xi)}{sh\zeta}\gamma_1\right] \quad (11.4)$$

$$\frac{1}{3-\delta_2^\beta}\left.\frac{\partial w^\beta}{\partial z}\right|_{z=1} = 2\xi\left[(\xi + \zeta cth\zeta l)\gamma_2 - \frac{\zeta\exp(-\xi l)}{sh\zeta l}\gamma_3\right] \quad (11.5)$$

$$\frac{1}{3-\delta_3^\beta}\left.\frac{\partial w^\beta}{\partial z}\right|_{z=1+l} = 2\xi\left[(\xi - \zeta cth\zeta l)\gamma_3 + \frac{\zeta\exp(\xi l)}{sh\zeta l}\gamma_2\right] \quad (11.6)$$

Substituting these values into (5.42)-(5.44) we obtain the system of three ordinary differential equations for $\gamma_1$, $\gamma_2$ and $\gamma_3$:

$$\frac{\partial \gamma_1}{\partial t} = \left[2\xi\zeta cth\zeta - 2\xi^2 - \eta\right]\gamma_1 - \frac{2\xi\zeta}{sh\zeta}\exp(-\xi)\gamma_2 \tag{11.7}$$

$$(\rho - 1)\frac{\partial \gamma_2}{\partial t} = -\frac{2\xi\zeta}{sh\zeta}\exp(\xi)\gamma_1 + P\gamma_2 - \frac{2\rho\xi\zeta}{sh\zeta l}\exp(-\xi l)\gamma_3 \tag{11.8}$$

$$\frac{\partial \gamma_3}{\partial t} = \frac{2\xi\zeta}{sh\zeta l}\exp(\xi l)\gamma_2 - \left[2\xi\zeta cth\zeta l + 2\xi^2 + \eta\right]\gamma_3 \tag{11.9}$$

where we have denoted for brevity

$$P = \rho\left[2\xi\zeta cth\zeta - 2\xi^2 - \eta\right] + \left[2\xi\zeta cth\zeta l + 2\xi^2 + \eta\right] \tag{11.10}$$

Solution of this system is obtained in the same way as of the system(10.4)-(10.5): substitution of $\gamma_1 = A\exp(\sigma t)$, $\gamma_2 = B\exp(\sigma t)$, and $\gamma_3 = C\exp(\sigma t)$ yields a linear homogeneous algebraic system for $A$, $B$ and $C$

$$G_{11}A + G_{12}B = 0 \tag{11.11}$$
$$G_{21}A + G_{22}B + G_{23}C = 0 \tag{11.12}$$
$$G_{32}B + G_{33}C = 0 \tag{11.13}$$

Here

$$G_{11} = 2\xi\zeta cth\zeta - 2\xi^2 - \eta - \sigma \tag{11.14}$$

$$G_{12} = -\frac{2\xi\zeta}{sh\zeta}\exp(-\xi) \tag{11.15}$$

$$G_{21} = -\frac{2\xi\zeta}{sh\zeta}\exp(\xi) \tag{11.16}$$

$$G_{22} = P - (\rho - 1)\sigma =$$
$$= \rho\left[2\xi\zeta cth\zeta - 2\xi^2 - \eta - \sigma\right] + \left[2\xi\zeta cth\zeta l + 2\xi^2 + \eta + \sigma\right] \tag{11.17}$$

$$G_{23} = -\frac{2\rho\xi\zeta}{sh\zeta l}\exp(-\xi l) \tag{11.18}$$

$$G_{32} = \frac{2\xi\zeta}{sh\zeta l}\exp(\xi l) \tag{11.19}$$

$$G_{33} = -\left[2\xi\zeta cth\zeta l + 2\xi^2 + \eta + \sigma\right] \tag{11.20}$$

For solutions of this system to exist the determinant of this system should equal zero,

$$\det \begin{pmatrix} G_{11} & G_{12} & 0 \\ G_{21} & G_{22} & G_{23} \\ 0 & G_{32} & G_{33} \end{pmatrix} = 0 \quad (11.21)$$

After some algebra (11.21) is *exactly* reduced to the following equation for $\sigma$

$$\det(G_{ij}) = \left[(2\xi^2 + \eta + \sigma)^2 - (2\xi\zeta)^2\right] \times \\ \times \left[2\xi\zeta cth\zeta l + 2\rho\xi\zeta cth\zeta + (1-\rho)(2\xi^2 + \eta + \sigma)\right] = 0 \quad (11.22)$$

As it is explained in Section 5, the root given by

$$2\xi\zeta cth\zeta l + 2\rho\xi\zeta cth\zeta + (1-\rho)(2\xi^2 + \eta + \sigma) = 0 \quad (11.23)$$

is negative, so from the point of possible instability only the roots of

$$(2\xi^2 + \eta + \sigma)^2 - (2\xi\zeta)^2 = 0 \quad (11.24)$$

are of interest. Two roots of (11.24) are

$$\sigma_{1,2} = -(2\xi^2 + \eta) \pm 2\xi\zeta \quad (11.25)$$

The solution of the system (11.11)-(11.13) is

$$\gamma_1 = A_1 \exp(\sigma_1 t) + A_2 \exp(\sigma_2 t) + A_3 \exp(\sigma_3 t) \quad (11.26)$$

$$\gamma_2 = B_1 \exp(\sigma_1 t) + B_2 \exp(\sigma_2 t) + B_3 \exp(\sigma_3 t) \quad (11.27)$$

$$\gamma_3 = C_1 \exp(\sigma_1 t) + C_2 \exp(\sigma_2 t) + C_3 \exp(\sigma_3 t) \quad (11.28)$$

The margin of stability corresponds to zero value of the largest root, $\sigma_1 = 0$. In its turn this means $\eta = 2\xi k$. The coefficients $A_n$, $B_n$, and $C_n$ are calculated in a standard way using initial values of $\gamma_i(0,k)$. Setting $t=0$ in (11.26)-(11.28) yields

$$A_1 + A_2 + A_3 = \gamma_1(0,k) \tag{11.29}$$
$$B_1 + B_2 + B_3 = \gamma_2(0,k) \tag{11.30}$$
$$C_1 + C_2 + C_3 = \gamma_3(0,k) \tag{11.31}$$

Additionally, for each root $\sigma_n$ equations (11.11) and (11.13) give a link between $A_n$, $B_n$ and $C_n$:

$$G_{11}(\eta, \sigma_n) A_n + G_{12}(\eta) B_n = 0 \tag{11.32}$$
$$G_{32}(\eta) B_n + G_{33}(\eta, \sigma_n) C_n = 0 \tag{11.33}$$

We are looking for the amplitudes at the onset of instability, so we need only "marginal" values of $G_{ij}$, that is

$$M_{ij}^n = G_{ij}(\eta, \sigma_n)\big|_{\eta=2\xi k} \tag{11.34}$$

For off-diagonal elements which don't depend on $\sigma_n$ we drop the upper index. To calculate the diagonal elements we need the "marginal" values of $\sigma_n$:

$$\sigma_1 = 0; \quad \sigma_2 = -4\xi(\xi+k); \tag{11.35}$$

$$\sigma_2 = -2\xi(\xi+k)\left[1 + \frac{\rho cth(\xi+k) + cth(l(\xi+k))}{1-\rho}\right] \tag{11.36}$$

Calculation of $M_{ij}^n$ yields

$$M_{11}^n = 2\xi(\xi+k)\left[cth(\xi+k) - 1\right] - \sigma_n \tag{11.37}$$
$$M_{33}^n = -2\xi(\xi+k)\left[cth(l(\xi+k)) + 1\right] - \sigma_n \tag{11.38}$$

$$M_{12} = -2\xi(\xi + k)\frac{\exp(-\xi)}{sh(\xi + k)} \quad (11.39)$$

$$M_{32} = 2\xi(\xi + k)\frac{\exp(l\xi)}{sh(l(\xi + k))} \quad (11.40)$$

From (11.32)-(11.33) we obtain

$$B_n = -\frac{M_{11}^n}{M_{12}}A_n; \quad C_n = \frac{M_{11}^n M_{32}}{M_{12} M_{33}^n}A_n \quad (11.41)$$

To compare the amplitudes at the onset of instability we need only $A_1$, $B_1$ and $C_1$. Using (11.29)-(11.31) and (11.41) we finally get

$$A_1 = \left[(M_{11}^3 - M_{11}^1)\left(1 - \frac{M_{11}^1 M_{33}^2}{M_{11}^2 M_{33}^1}\right)\right]^{-1} \times$$
$$\times \left[\gamma_{10}M_{11}^3 + \gamma_{20}M_{12}\left(1 - \frac{M_{33}^2}{M_{11}^2}\right) - \gamma_{30}\frac{M_{12}M_{33}^3 M_{33}^2}{M_{32}M_{11}^2}\right] \quad (11.42)$$

as well as corresponding expressions for $B_1$ and $C_1$. For large $k$ the approximate expressions for these amplitudes become:

$$A_1 \simeq \gamma_1(0,k) - (1-\rho)\exp(-2\xi - k)\gamma_2(0,k) - $$
$$-\rho\exp(-(2+l)\xi - k)\gamma_3(0,k) \quad (11.43)$$

$$B_1 \simeq \exp(-k)\gamma_1(0,k) - (1-\rho)\exp(-2(\xi + k))\gamma_2(0,k) - $$
$$-\rho\exp(-(2+l)\xi - 2k)\gamma_3(0,k) \quad (11.44)$$

$$C_1 \simeq \exp(-k(1+l))\gamma_1(0,k) - (1-\rho)\exp(-2\xi - k(2+l))\gamma_2(0,k) - $$
$$-\rho\exp(-(2+l)\xi - k(2+l))\gamma_3(0,k) \quad (11.45)$$

*Petro Mchedlov-Petrosyan*

*A.I.Akhiezer Institute for Theoretical Physics,*

*National Science Center "Kharkov Institite of Physics & Technology",*

*1, Akademicheskaya Str., Kharkov, Ukraine 61108*

peter.mchedlov@free.fr

T. +(38) 057 704 02 60

Manfred Martin

Institute of Physical Chemistry

RWTH Aachen University

Landoltweg 2

52056 Aachen, Germany

martin@rwth-aachen.de

T. +49-241-80-94712